\documentclass[onecolumn,floatfix,preprintnumbers,amsmath,amssymb,prb]{revtex4}
\usepackage{graphicx}% Include figure files
\usepackage{dcolumn}
\usepackage{latexsym}
\usepackage{epsfig,color}
\usepackage{epstopdf}
\usepackage{epstopdf}
\usepackage{amsmath}
\usepackage{amsfonts}
\usepackage{amssymb}
\usepackage{float}
\usepackage{multirow}

\newcommand{\be}{\begin{equation}}
\newcommand{\ee}{\end{equation}}
\newcommand{\bea}{\begin{eqnarray}}
\newcommand{\eea}{\end{eqnarray}}

\newcommand{\p}{\partial}
\newcommand{\s}{\sigma}

\newcommand{\la}{\langle}
\newcommand{\ra}{\rangle}
\newcommand{\rd}{\mbox{d}}
\newcommand{\ri}{\mbox{i}}

\newcommand{\nn}{\nonumber}

\newcommand{\vare}{\varepsilon}

\newcommand{\vxi}{\mbox{\boldmath $\xi$}}

\begin{document}

\input{epsf}

\title{Classical impurities and boundary Majorana zero modes
in quantum chains}

\author{Markus M\"uller$^{1,2,3}$}
\author{Alexander Nersesyan$^{1,4,5}$}

\affiliation{
$^1$The Abdus Salam International Centre for Theoretical Physics, 34151, Trieste, Italy \\
$^2$Condensed matter theory group, Paul Scherrer Institute, CH-5232 Villigen PSI, Switzerland\\
$^3$Department of Physics, University of Basel, Klingelbergstrasse 82, CH-4056 Basel, Switzerland\\
$^4$ ITP, Ilia State University, 0162, Tbilisi, Georgia\\
$^5$ Andronikashvili Institute of Physics, 0177, Tbilisi, Georgia
}

\begin{abstract}

We study the response of classical impurities in quantum Ising chains. The 
{$\mathbb{Z}_2$ degeneracy they entail renders the existence of two decoupled Majorana modes at zero energy an exact property of a finite system at arbitrary values of its bulk parameters.
We trace the evolution of these %Majorana zero 
modes across the transition from the disordered phase to the ordered one and analyze the 
concomitant
qualitative changes of local magnetic properties of an isolated impurity.
In the disordered phase, the two ground states differ only close to the impurity, and they are related by the action of an explicitly constructed quasi-local operator. In this phase the {local}
transverse spin susceptibility %of the impurity 
follows a Curie law.} 
The critical response of a boundary impurity
is logarithmically divergent and maps to the two-channel Kondo problem, while it
saturates for critical bulk impurities, as well as in the ordered phase. The results for
the Ising chain translate to the related problem of a resonant level coupled to a 1d
p-wave superconductor or a Peierls chain, whereby the magnetic order is mapped to
topological order.  We find that the topological phase always exhibits  a continuous
impurity response to local fields as a result of the level repulsion of local levels
from the boundary  Majorana zero mode. In contrast, the disordered phase generically
features a discontinuous magnetization or charging response. This difference
constitutes a general thermodynamic fingerprint of topological order in phases with a bulk gap.
\end{abstract}

\maketitle

%%%%%%%%%%%%%%%

\section{Introduction}

In recent years there was a substantial boost in the search for signatures of Majorana zero modes (MZM) that may emerge as localized quasiparticles in various condensed matter realizations, because of their potential for quantum computation \cite{dassarma-rev, alicea1}.
MZMs occuring at boundaries or defects (domain walls and vortices) in low-dimensional topological superconductors are of particular importance because of their non-Abelian anyonic statistics that uncovers new prospects for storage and manipulation of quantum information \cite{kitaev1,nayak}.

\medskip

In his seminal paper Kitaev \cite{kitaev2} proposed a one-dimensional 
model of a spinless $p$-wave superconductor (1DPS) 
\bea
 H = - \mu \sum_{n=1}^N (a^{\dagger}_n a_n - 1/2)
 + \frac{1}{2}
\sum_{n=1}^{N-1} \left( t a^{\dagger}_n a_{n+1}  + \Delta a^{\dagger}_n a^{\dagger}_{n+1} + h.c.\right)
\label{p-1D-ham}
\eea
(the usual negative sign of the hopping term can be obtained by the transformation $a_n\to (-1)^n a_n$, which changes the signs of $t$ and $\Delta$).
This model has a topologically
non-trivial massive phase that supports
localized Majorana modes 
at the ends of the chain. For a macroscopically large system 
these boundary modes can be regarded as unpaired,
in which case they represent 
a non-local realization of a doubly degenerate fermionic zero-energy state.
The spatial separation of the two MZMs ensures the immunity 
of the topologically degenerate ground state of the 1DPS against weak local perturbations {(as long as quasi-particle poisoning can be neglected, and thus fermion parity is conserved)}, 
making such a system potentially useful for the needs of quantum
computation. Thus it is of great theoretical interest and practical importance to identify the physical properties of the edge of such a 1D system that 
can serve as evidence for the existence
of boundary MZMs.
\medskip

It has soon been realized that principal features of the Kitaev 1D model \cite{kitaev2}
can be reproduced experimentally using a quantum wire with a strong spin-orbit coupling
in the presence of an external magnetic field and the proximity effect
with a conventional s-wave superconducting substrate \cite{lutchyn,oreg}. Much theoretical and experimental
effort is currently going into finding unambiguous signatures of MZMs in various set-ups.
Important steps forward in this direction include tunneling
spectroscopy experiments\cite{mourik,mag}, whose findings,
{in particular, the zero-bias conductance peak observed in one-dimensional semiconductor-superconductor
contacts,}
were
consistent with theoretical predictions (see Ref.~\onlinecite{dassarma-rev} for a recent review).
\medskip

Closely related to the 1DPS model is the quantum  Ising chain (QIC), described
by the Hamiltonian:
\be
H = - J \sum_{n=1}^{N-1} \s^x _n \s^x _{n+1} - h \sum_{n=1}^N \s^z _n. \label{ham_QIC}
\ee
Here $\s^{\alpha}_n$ are Pauli matrices, $J>0$ is the exchange interaction and $h$
is a transverse magnetic field which endows the spins with quantum dynamics. 
The model %(\ref{ham_QIC}) 
possesses a $\mathbb{Z}_2$-symmetry associated with the global
transformation  $P_S \s^x _n P^{-1}_S = - \s^x_n$, where %induced by the operator
$P_S = \prod_{j=1}^N \s^z _j$, $~[H,P_S]=0$. 
This is an exactly solvable 
quantum 1D model which, by virtue of the transfer matrix formalism, is related 
to the classical 2D Ising model \cite{schultz,kogut,mussardo}.
The Jordan-Wigner (JW) transformation maps the many-body problem (\ref{ham_QIC}) %QIC 
onto a quadratic
model of spinless fermions, the latter actually being 
a particular realization of the 1DPS (\ref{p-1D-ham}) with a fine-tuned pairing amplitude
$\Delta = \pm t$.
Close to criticality, in the field-theoretical limit, the QIC represents a (1+1)-dimensional theory of a massive
Majorana fermion \cite{mussardo}.
The topological phase of the 1DPS
corresponds to
the ordered phase of the QIC, which (in the thermodynamic limit) is characterized by spontaneously broken symmetry and a two-fold degeneracy of the ground state (up to an exponentially small splitting). The ordered phase ($J>h$) is separated from the 
topologically trivial, disordered phase ($J<h$) by a quantum critical point ($J=h$).
As follows from the Kramers-Wannier
duality\cite{kogut,mussardo}, the two massive phases of the QIC (\ref{ham_QIC}) have identical bulk spectrum; however, they differ in the boundary
conditions at the edges of a finite chain, %\cite{kitaev2,kitaev3}, 
reflecting
their topological distinction. 
\medskip

This difference is clearly seen from the 
Kitaev-Majorana (KM) representation 
of the QIC  \cite{kitaev2, kitaev3}: the $N$-site spin chain (\ref{ham_QIC}) 
is equivalent to a $2N$-site tight-binding model of real (Majorana) fermions with nearest-neighbor
couplings, as will be briefly reviewed below:
\bea
H = \ri \sum_{j=1}^N \left( h c_{2j-1} c_{2j} + J c_{2j} c_{2j+1} \right), ~~~
c^{\dagger}_i = c_i, ~~~\{ c_i, c_j \} = 2 \delta_{ij}. \label{ham-c}
\eea
In the representative limits $h/J \to 0$ and $J/h \to 0$ a greatly simplified qualitative picture
emerges. 
For $h \to 0$ one finds
two decoupled boundary MZMs, $c_1$ and $c_{2N}$, in the otherwise dimerized chain,
implying  a two-fold degeneracy of the ground state, while for $J\to 0$
the KM lattice has a full dimer covering and the ground state is unique.
At finite $h/J < 1$ the exact degeneracy between the two boundary MZMs in the ordered phase
is removed.
In spin language, the level splitting is caused by quantum tunneling between the two classical
Ising vacua caused by the propagation of a magnetization kink from one end of the chain to the other.
For a macroscopically long chain,  the tunnelling amplitude is exponentially small,
$t_{\rm eff} \sim J \exp(-L/\xi)$ ($L$ and $\xi$ being the length of the chain and
the correlation length, respectively). However, within this accuracy the two boundary
Majorana modes remain true zero modes, and their existence implies the two-fold degeneracy of the ground state in
the ordered phase of the QIC.
\medskip

It is worth mentioning that the usefulness of the Majorana fermions in the QIC hinges on the 
exact $\mathbb{Z}_2$ symmetry of the spin model (\ref{ham_QIC}). Local terms (in the $\s^x$ or $\s^y$)
which break the Ising symmetry induce non-local couplings among the Majoranas and spoil the degeneracy
of the edge modes. In contrast, the topological phase of the 1DPS is robust against any local perturbations in the fermionic basis. Nevertheless, within the symmetry protected sector, the QIC exhibits very similar physical properties as the 1DPS.
\medskip

In this paper we aim at identifying clear physical differences associated with the presence or absence of Majorana edges modes in the topologically ordered or non-ordered phases, respectively.  We focus on the effects of an impurity that interrupts an otherwise homogeneous 1d chain, or terminates it. We study its spectral weight and its response to locally applied external fields. Our main result is the finding summarized in Table~\ref{table:MainIllustration}: the local susceptibility of such an impurity can serve as a probe for Majorana edges modes in the adjacent bulk phase(s). In particular, we find that the non-topological phase is characterized by a discontinuity in the polarization response  of the impurity to an external field, and a concomitant divergence of the susceptibility. In contrast, topological order and the associated MZMs 
quench such a divergence. This effect may serve as a {\em thermodynamic, equilibrium} tool in the search for direct traces of Majorana edge modes, which so far have been sought mostly in transport properties at zero bias.
\medskip

\begin{table}[h]
\begin{center}
 \begin{tabular}{ | l || c | c | }
   \hline 
     & Quantum Ising chain & 1d p-wave superconductor \\ \hline \hline
    \multirow{2}{*}{\parbox[t]{3.5cm}{Disordered phase}} & \raisebox{-0\height}{\includegraphics[width=2in]{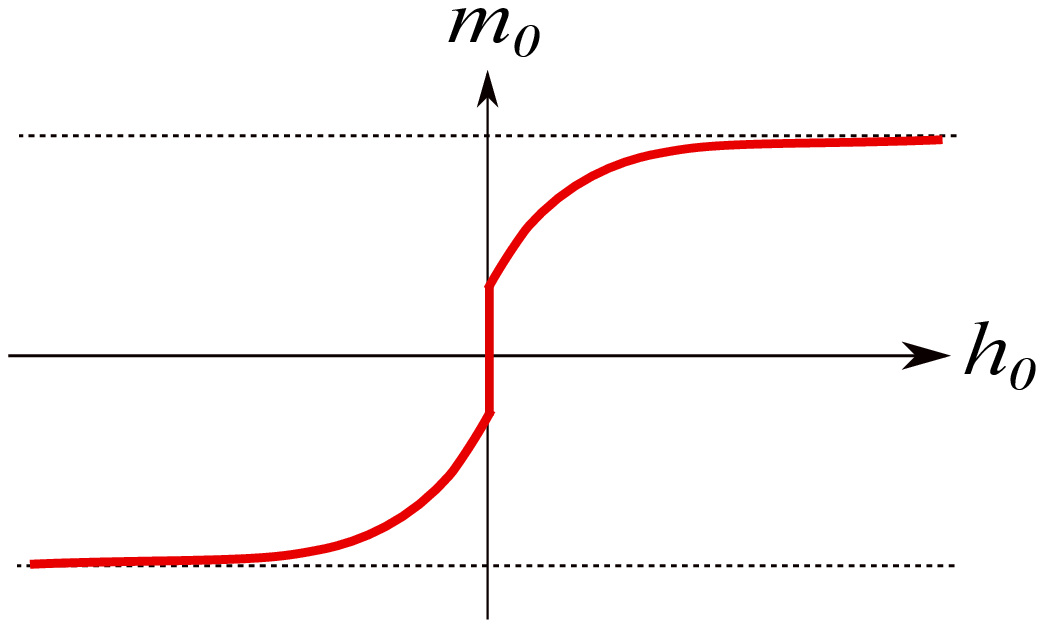}
   } & \raisebox{-0\height}{\includegraphics[width=2in]{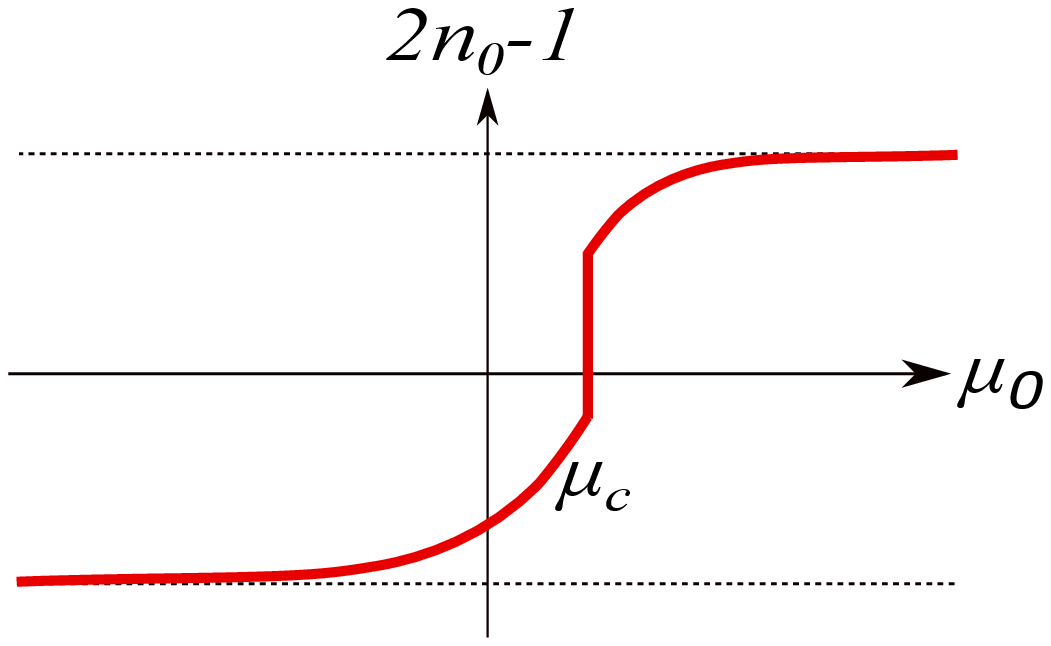}} \\ 
   & \raisebox{-.5\height}{\includegraphics[width=2in]{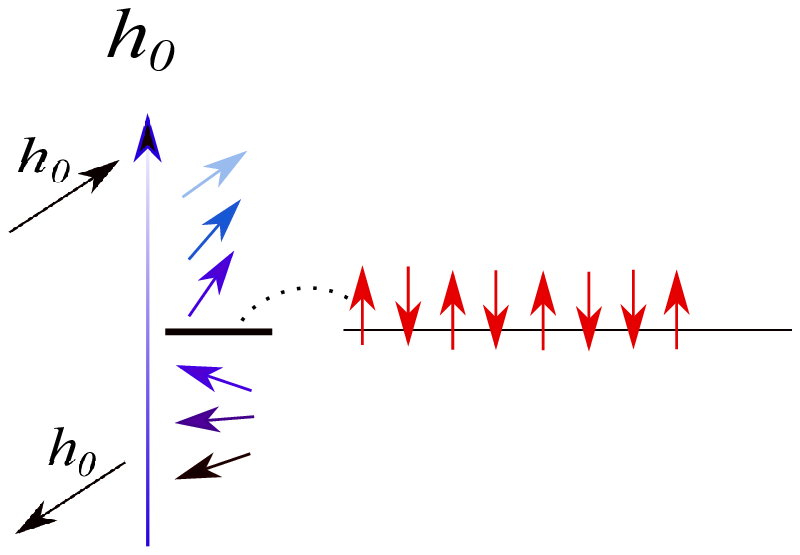}} 
   & \raisebox{-.5\height}{\includegraphics[width=2in]{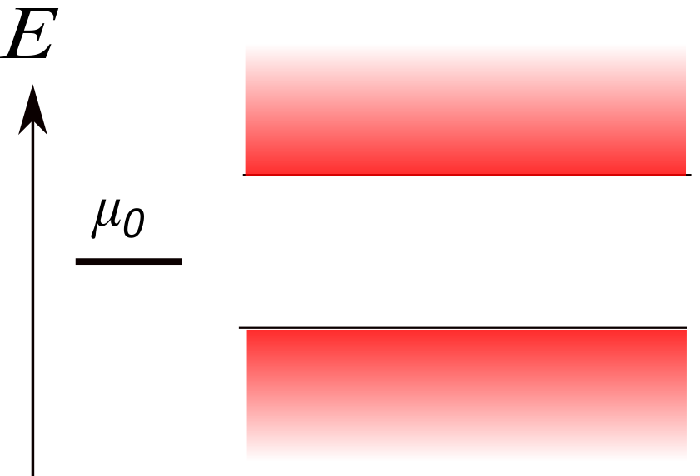}}\\ 
   \hline
    \multirow{2}{*}{\parbox[t]{3.5cm}{Ordered phase \\ (magnetic/topological)} } & \raisebox{-0\height}{\includegraphics[width=2in]{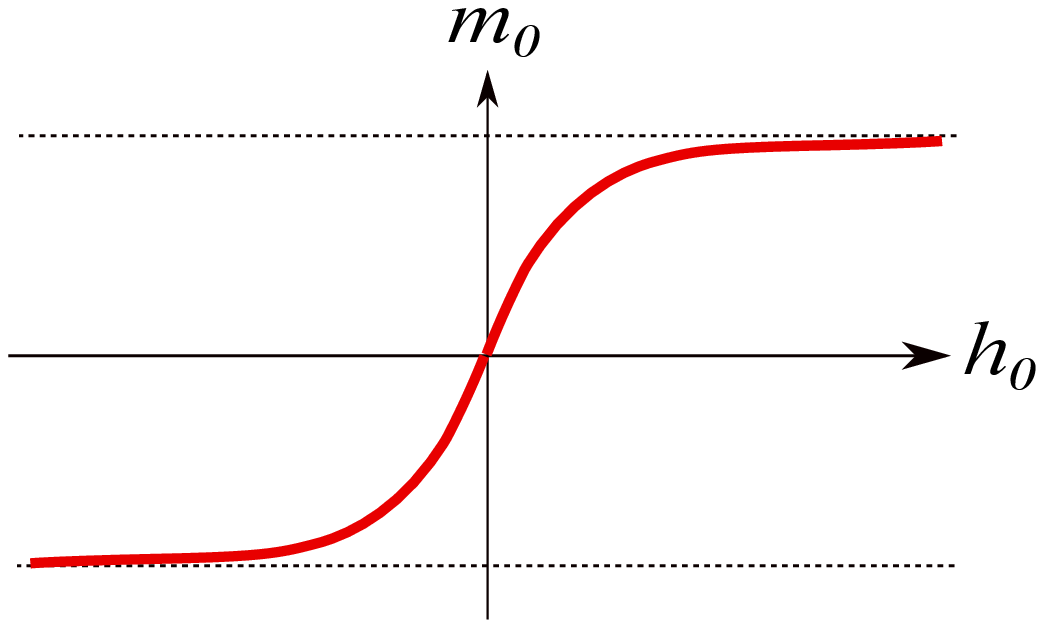}} & \raisebox{-0\height}{\includegraphics[width=2in]{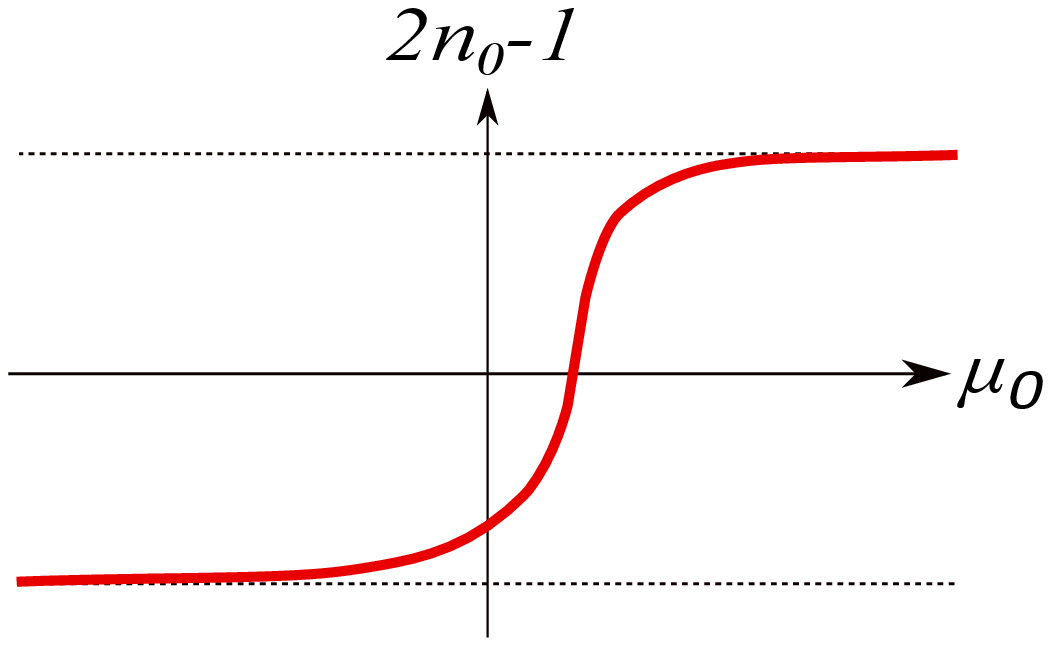}} \\ 
    & \raisebox{-.5\height}{\includegraphics[width=2in]{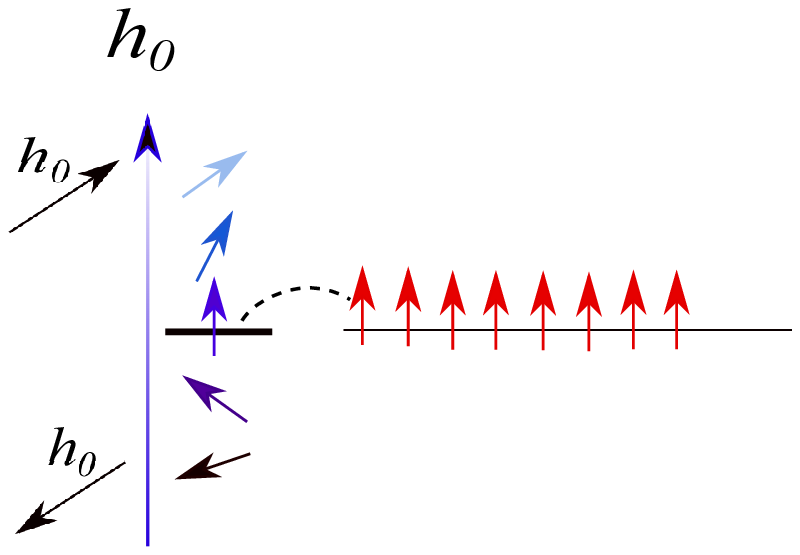}
   } & \raisebox{-.5\height}{\includegraphics[width=2in]{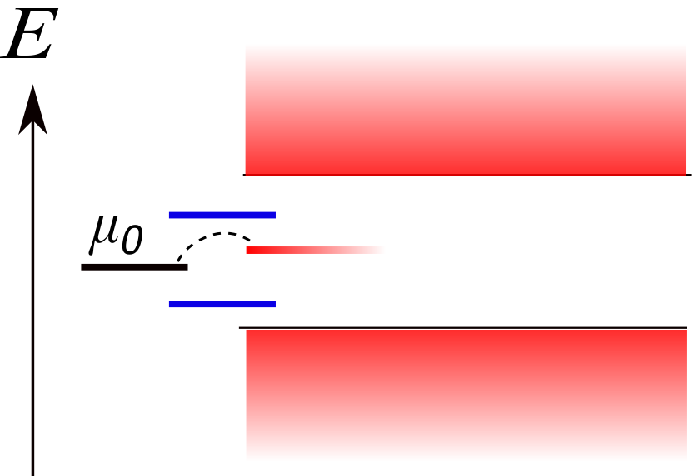}}\\
   \hline
 \end{tabular}\end{center}
 \caption{{The response of an impurity to local fields reveals the nature of the chain that it  couples to. Ising spin chains and spinless p-wave superconductors are closely analogous. At $T=0$, the spin $\sigma_0$ undergoes a discontinuous flip at $h_0=0$ in the paramagnetic phase, where a dressed free spin remains localized at the  edge. On the ferromagnetic side, the spontaneously broken symmetry in the ground state generates a longitudinal field on the impurity and renders the transverse magnetization response $m_0(h_0)$ smooth.
 Upon JW transformation the Ising system maps to a single-level quantum dot which couples to a 1d p-wave superconductor. Thereby the dot's occupation $n_0$ as a function of the local potential $\mu_0$ takes the role of $m_0(h_0)$. In the topologically trivial phase the  level localized on the quantum dot can be driven through zero energy, inducing a discontinuous jump in $n_0(\mu_0)$. In this case, generic local couplings will shift the jump to arbitrary $\mu_c$. In contrast, in the topological phase the coupling to the boundary Majorana zero mode  repels the energy of the localized boundary state away from zero and thus renders $n_0(\mu_0)$ smooth. 
 This paper analyzes the impurity response in the various phases, especially  close to the degeneracy point ($h_0=0$ or $\mu_0=\mu_c$). The susceptibility has Curie-like divergence in the disordered phase, while it saturates in the ordered phase. At criticality, the problem maps to the 2-channel Kondo effect, and accordingly, the susceptibility is a logarithmically diverging function of temperature.
 }}
\label{table:MainIllustration}
\end{table}

For illustrative purposes and analytical convenience we focus on  %a finite-size 
the QIC model  in which the presence of impurities of a certain kind 
enforces all energy levels of the system to be two-fold
degenerate even for a finite system. 
The impurities we have in mind represent lattice sites where the local transverse magnetic
field vanishes. The spins residing at these sites
are unable to flip and, therefore, are classical.
The $\mathbb{Z}_2$ degeneracy of the ground state 
makes the existence of two decoupled Majorana modes at zero energy
an exact property of a \emph{finite} system at \emph{arbitrary} values of the bulk 
parameters $J$ and $h$. The goal of this work is
to trace %understand 
the evolution of the associated MZMs across 
the transition from the disordered phase to the ordered one and describe
the corresponding
qualitative changes of  measurable quantitites -- the
spectral weight (density of states) of the impurity
spin and the local magnetic susceptibility defined as the linear response
to a small transverse magnetic field. For a 1D p-wave superconductor the
equivalent quantities are the average occupancy of the impurity fermionic level
and the local charge susceptibility.
%\AN{ to be continued} 
\medskip

The paper is organized as follows.
In Sec.~\ref{MKrep} we briefly overview the QIC in the KM representation, which in
Sec.~\ref{qual}  is used to qualitatively describe the main
features of a quantum Ising chain containing classical-spin impurities: the presence
of a free local spin with a local Curie susceptibility  in the disordered phase, its
delocalization at the phase transition and the transformation of the spectral
degeneracy from locally differing groundstates to globally differing Ising 
symmetry-broken states.
\medskip

In Sec.~\ref{sec:conservedspin} we explicitly construct a triplet of  conserved
operators which obey the standard spin $1/2$ algebra, and are quasi-local in the
disordered phase. We relate their existence  to the integrable character of the
considered models, and compare with similar conserved operators in many-body
localized systems.
\medskip

In Sec.~\ref{reduction-0} we consider a single classical impurity in an Ising
chain close to criticality. Taking the scaling limit, we establish the connection
with {massive versions of} previously studied resonant-level models 
{where the impurity couples to 
two channels
%channels 1 or 2 
of Majorana
fermions}, and we explain how to compute the observables of interest using the
Green's functions of the auxiliary Majorana fermions. Sec.~\ref{semi}
contains our central results for a boundary impurity. We evaluate the impurity
spectral weight in both phases and calculate the temperature dependence of the
transverse susceptibility of the impurity spin finding a rich behavior across the
quantum critical window. At criticality, {the QIC with a boundary 
impurity coincides with the Majorana resonant-level model discussed earlier
by Emery and Kivelson\cite{EK} in their studies of the two-channel Kondo problem.
In this regime the impurity spin has a logarithmically divergent
low-temperature susceptibility.} %local spin susceptibility
%spin behaves like a two-channel
%Kondo problem, with a logarithmicaly diverging low $T$ susceptibility. 
This is
intermediate between the Curie asymptotics of the disordered phase and the
saturating susceptibility in the ordered phase. Sec.~\ref{1DpwS} establishes the
connection with the one-dimensional p-wave superconductors to which the QIC maps
under JW transformation. In particular, we find that the local
compressibility of an impurity site (a quantum dot coupled to a superconducting
wire) provides a thermodynamic signature of the presence or absence of topological
order in the superconductor: The topological phase with its boundary  Majorana zero
mode forces the charge occupation of the quantum dot to be a smooth function of
local potential acting on the dot. This is in contrast to the topologically trivial
superconducting phase of the wire, in the presence of which the occupation of the
dot generically undergoes discrete jumps as a function of applied gate voltage.
Sec.~\ref{bulk} analyzes an impurity in the bulk and summarizes the salient
features of the susceptibility, and  how it differs from a boundary impurity. The
symmetrically coupled impurity is shown to map to a semi-infinite Peierls chain
coupled to a boundary impurity.  The concluding section~\ref{summary}
summarizes the main results, and discusses how generally topological order in 1
dimension may be detected by the absence of discontinuous response to local fields
acting on impurities.

\section{Quantum Ising Chain  in Kitaev-Majorana representation}\label{MKrep}
%%%%%%%%%%%%%%%%%%%%%%%%%%%%%%%%%%%%%%%%%%%%

We start our discussion with a brief
overview of the KM representation of the QIC \cite{kitaev2,kitaev3}.
The non-local JW transformation %that 
expresses the lattice spin-1/2 operators
$S^{\alpha}_n = (1/2) \s^{\alpha}_n$ in terms of spinless fermionic operators
$a_n$ and $a^{\dagger}_n$:
\bea
\s^+ _n = \s^x + \ri \s^y = 2(-1)^n a^{\dagger}_n \exp \left( \ri \pi \sum_{j=1}^{n-1}
a^{\dagger}_j a_j  \right), ~~~
\s^z _n = 2a^{\dagger}_n a_n - 1.
\label{jw}
\eea
The Hamiltonian (\ref{ham_QIC}) then transforms to a quadratic form
\bea
H &=& \sum_n
[ J (a^{\dagger}_n - a_n)(a^{\dagger}_{n+1} + a_{n+1})
- h (a^{\dagger}_n - a_n)(a^{\dagger}_{n} + a_{n})]
\nn\\
&=& J \sum_n \left( a^{\dagger}_n a_{n+1} + h.c.  \right)
+ J \sum_n \left( a^{\dagger}_n a^{\dagger}_{n+1} + h.c.  \right)
- 2h \sum_n \left(a^{\dagger}_n a_n - 1/2 \right) \label{ham-a}.
\eea
It does not conserve the particle number $N_F = \sum_i a^{\dagger}_i a_i$, but only the fermionic number parity: $P_S = (-1)^{N_F}$. The latter is simply the fermionic expression for the Ising flip operator.
Notice that 
(\ref{ham-a}) is a 1DPS model (\ref{p-1D-ham})
with a fine-tuned amplitude of the Cooper pairing $\Delta$ equaling the hopping strength $t$.
\begin{figure}[hbbp]
\centering
\includegraphics[width=2.3in]{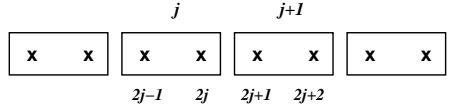}
\caption{\footnotesize Kitaev-Majorana  chain with twice as many lattice sites as the Ising spin chain. Boxes indicate the Ising spin degrees of freedom which are split into two Majoranas.}
\label{c-lattice}
\end{figure}
\noindent
The underlying Majorana structure of the Hamiltonian (\ref{ham-a}) is manifest. 
A physical site $j$ of the original lattice, shown as a box 
in Fig.~\ref{c-lattice},  is associated with
a local Fock space of the complex fermion $(a_j, a^{\dagger}_j)$.
Each physical site can then be split into a pair
of "Majorana" sites, shown by crosses in Fig.~\ref{c-lattice}, where real fermion
operators $\{ c_j \}$
are defined,
\bea
c_{2j-1} = a^{\dagger}_j + a_j, ~~c_{2j} =- \ri (a^{\dagger}_j - a_j), ~~
\{ c_j, c_l \} = 2 \delta_{jl}. \label{c_ops}
\eea
The Hamiltonian (\ref{ham-a}) then transforms into the $2N$-site
KM lattice model (\ref{ham-c}).
At $J \neq h$ the translational invariance of the KM lattice (\ref{ham-c})
is broken, entailing
a spectral gap.
For later purposes, it is crucial to keep in  mind that
the transverse field $h$ tends to pair the  $c$-fermions belonging to the same
physical site, while the exchange interaction $J$ couples fermions belonging to 
neighboring boxes. 
\medskip

The Kramers-Wannier duality~\cite{kogut,mussardo} transforms 
  the original set of spin operators $\s^{\alpha}_n$ to the so-called
disorder operators $\mu^{\alpha}_{n}$ associated with the links $<n,n+1>$, 
\be
\mu^x _{n} = \prod_{j=1}^{n-1} \s^z _j, ~~~~
\mu^x _n \mu^x _{n+1} = \s^z_n, ~~~~ \mu^z _n = \s^x _n \s^x _{n+1}.
\label{duality}
\ee
Thereby it maps the Hamiltonian (\ref{ham_QIC}) into the same model, but with  $J$ and $h$ interchanged, up to boundary terms.
In the KM representation (\ref{ham-c}), the non-local 
transformation (\ref{duality}) simply reduces to a translation by one lattice spacing.

\bigskip

%%%%%%%%%%%

%%%%%%%
\section{Classical impurity spins in quantum Ising chain: Qualitative picture}\label{qual}
%%%%%%%%%%%%%%%%%%%

%%%%%%%
Consider the ordered phase of an inhomogeneous QIC with
locally varying transverse magnetic fields $h_n$.
%allowing local fluctuations of
%the transverse magnetic field.
Imagine that a magnetization kink, separating two classical Ising vacua with opposite
spin polarizations,
travels along the chain from its left end to the right one,
with each elementary step being associated with a spin reversal caused %being induced 
by a nonzero local 
field $h_n$. For $h_n \ll J$ the vacuum-vacuum tunneling amplitude is proportional to
(one can always assume that $h_n \geq 0$)  
\be
t_{\rm eff} {\approx} J \prod_{n=1}^N \left(\frac{h_n}{J}\right).\label{Delta-inhom}
\ee
Therefore,
if at some lattice site the local field vanishes,
$t_{\rm eff}$ vanishes as well implying that the boundary Majorana modes become 
true degenerate zero modes, \emph{even for finite N}. Physically, this %result 
follows from the fact that at the impurity site
spin reversal is impossible. 
The spin localized at such a site is classical, i.e., unable to flip. 
For a kink moving along the chain the zero-field site represents an infinitely high
barrier which blocks its further propagation. Mixing of the boundary Majorana states located
at the opposite boundaries  thus becomes impossible;  hence an %the 
exact $\mathbb{Z}_2$ degeneracy of the ground state.

\medskip

Let us make this statement more precise.
Consider 
a QIC with a zero-$h$ impurity, say at $n = 0$:
\be
H = - J \sum_n \s^x _n \s^x _{n+1} - h \sum_{n\neq 0}\s^z _n \label{ham+imp}.
\ee
There are two 
operators, $\s^x _0$ and $P_S = 
\prod_{n=1}^N \s^z _n$, which
commute with the Hamiltonian, but anticommute with each other. Each of these operators squares to unity.
If  $\psi_{\s}$ is an eigenstate of both $H$ and $\s^x _0$,
\[
H \psi_{\s} = E \psi_{\s}, ~~~\s^x _0 \psi_{\s} = \s \psi_{\s}
~~(\s = \pm 1)
\] 
then the anticommutation relation $\{ \s^x _0 , P_S \} = 0$ implies that
the state $\psi' _{\s'} = P_S \psi_{\s}$ is also an eigenstate of $H$ with the same energy $E$
but with $\s' = - \s$.  
Hence all energy levels of the system %impurity QIC model
are two-fold degenerate
at \emph{arbitrary} $J/h$.
This conclusion remains obviously valid for an arbitrary number $N_i$ of zero-field impurities
in which case the Hamiltonian has the form:
\be
H = - J \sum_{n} \s^x_n \s^x _{n+1} - h \sum_{n \notin {\cal J}} \s^z _n,
\ee
where ${\cal J}$ is the set of all impurity sites.
Obviously, $H$ commutes with  $P_S$  and the $N_i$ operators $\s^x _m ~(m \in {\cal J})$.
The  operators $\s^x _m $ commute among each other, whereas all of them anticommute with $P_S$. 
Since 
$P_S$ simultaneously inverts the signs of the eigenvalues of all impurity spins
$\s^x _m$, by the same argument as before  one concludes
that the energy levels of the many-impurity model are also two-fold degenerate (but generically there is no higher degeneracy). 
Thus,
irrespective of the nature of the bulk phase, ordered or disordered,
the ground state of a QIC with classical impurity spins is 
$\mathbb{Z}_2$-degenerate.
\begin{figure}[hbbp]
\centering
\includegraphics[width=3.3in]{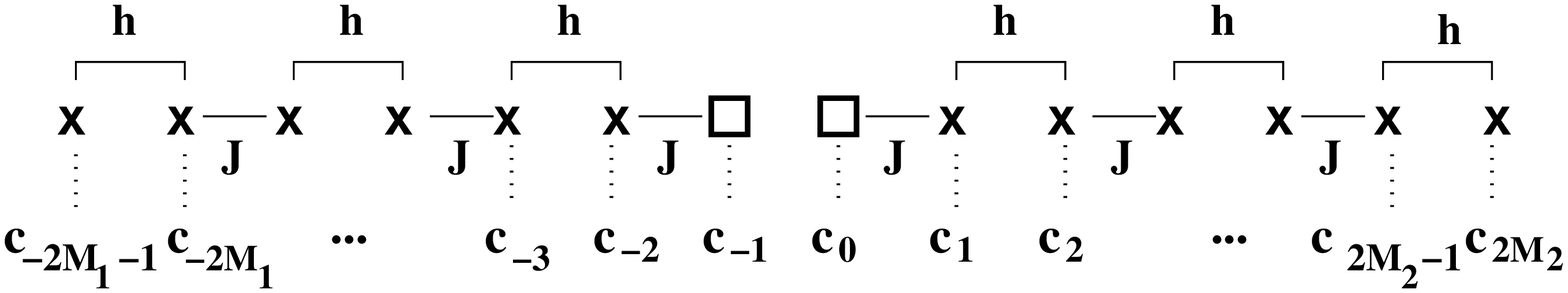}
\caption{\footnotesize KM representation of QIC with $h_j = h (1-\delta_{j0})$.}\label{kit}
\end{figure}
\medskip

To understand how this degeneracy is physically realized
in the disordered and ordered bulk phases, let us
look at the KM representation of a finite
QIC with a zero-field impurity at the origin. Let
$M_1$ and $M_2$ be the numbers of the
lattice sites to the left and to the right of the impurity site, 
the total number of physical sites being $N = M_1 + M_2 +1$. 
The pattern of pairings of neighboring $c$-operators on the corresponding KM lattice
is displayed in Fig.~\ref{kit}.
The operators $c_{-1}$ and $c_0$ shown by two squares originate from the impurity site
$n=0$. As opposed to all other $c$-operators, these two operators are unpaired,
$\la c_{-1} c_0 \ra = 0$,
because $h_0 = 0$. As a consequence, the the impurity cuts the KM chain in Fig.~\ref{kit} into two disconnected pieces containing $2M_1 + 1$ 
and $2M_2 + 1$ sites, respectively.
Since these numbers are odd, the emerging situation is
special because a piece of a Majorana lattice with an odd number of
sites does not represent a segment of the original physical lattice where the
spins $\s^{\alpha}_n$ are defined. 
\begin{figure}[hbbp]
\centering
\includegraphics[width=3.5in]{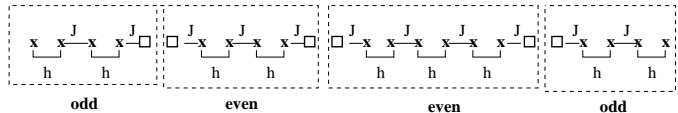}
\caption{\footnotesize QIC with several classical (zero-field) impurities in the 
KM representation}
\label{3imp}
\end{figure}
\medskip

Local fluctuations of the Hamiltonian parameters
can indeed cut the KM lattice into disconnected pieces. This 
can be due to vanishing local 
values of $J$ or $h$ at certain links or sites.
The former case is trivial: randomly distributed links with vanishing exchange
couplings cut the original spin chain into segments, each representing a smaller-size
chain with open boundaries.  %This implies that, 
On the KM lattice, each
disconnected segment contains an even number of sites.
This picture should be contrasted with the situation emerging in a QIC
containing sites with vanishing local magnetic fields.  An example 
is shown in Fig.~\ref{3imp}. 
For a spin chain with $N_{i}$ impurities,
the corresponding KM lattice decouples
into $N_{i} + 1$ pieces: $N_{i} - 1$ of them contain even numbers
of $c$-sites and two more pieces with odd numbers of sites are attached to the boundaries.
\medskip

Now, a finite KM
chain with an odd number of sites necessarily contains an exact MZM.
Indeed, the Hamiltonian of such a chain is a quadratic form
${\cal H} = \ri \sum_{jk} A_{jk} c_j c_k$, where $A_{jk}$ is a real, antisymmetric
$N\times N$ matrix, where $N$ is odd. Consequently,
${\rm det~}\hat{A} = 0$ and so the set of eigenvalues of $\hat{A}$
necessarily contains a zero eigenvalue. 
So a finite QIC with a classical-spin impurity should have
two exact MZMs.
To understand where these zero modes are located, 
it is instructive to turn again to
%let us %following Kitaev\cite{kitaev2}
%turn to 
the limiting cases $h=0, J\neq 0$ and $J=0, h\neq 0$.
Considering for instance the right segment of the KM chain
in Fig.~\ref{kit}, 
one finds that a KM
chain with an odd number of sites contains a zero mode at the right boundary if
$J>h$ or at the left boundary if $J<h$. For the left segment of the KM chain
the situation is just inverted.
The boundary MZM 
will have a finite localization radius at any $J \neq h$
and move from one boundary to the other as the critical point is crossed.
\medskip

We thus arrive at the following physical picture for 
%Consider a long sample of 
a QIC 
with an impurity spin in the bulk.  In the disordered phase 
the two MZMs are located close to the impurity site.
In this phase, the free impurity spin of the limit $J\to 0$ retains its identity, despite getting dressed and delocalizing over a finite length scale $\xi$:
the local transverse spin susceptibility, defined as the linear response to a small transverse field
$h_0$, follows a Curie law:
$
\chi_0 \sim T^{-1}.
$
The impurity zero modes
are fragile because they are not spatially separated:
application of a small transverse
local field $h_0$ will produce Zeeman splitting of the zero-energy levels and polarize
the impurity spin in the $z$-direction.
\medskip

Once the system passes over to the ordered phase, the exact
$\mathbb{Z}_2$ degeneracy becomes  a non-local property of the ground state.
The free impurity spin
"disappears" because it gets strongly coupled to the rest of the system by the 
classical Ising exchange. 
The local spin susceptibility $\chi_0$ is finite in the zero-temperature limit.
However, in agreement with the Kitaev's picture\cite{kitaev2}, 
 the disappearance of the impurity spin in the ordered phase is accompanied, in the fermionic language, by the appearance of two spatially
separated MZMs at the end-points of the chain.
In other words, these boundary zero modes represent what the spectral weight of the
local impurity spin transforms into when the system undergoes the quantum phase transition from
the disordered to the ordered phase. Let us stress again that these are \emph{exactly
degenerate} boundary MZMs in  a \emph{finite} quantum Ising chain -- an effect
caused by the zero-field impurity, which kills the tunneling between the Ising vacua.
\medskip

Consider now the case of
a finite number $N_i$ of classical-spin impurities.
As already explained, there should again be two exact MZMs in the ground state.
In the ordered phase, %sustains the presence of impurities, and 
the Kitaev's picture  \cite{kitaev2} of
two boundary MZMs
at the end-points of the chain is intact. In the disordered phase
there are $2N_{i}$ boundary Majorana modes localized in the vicinity of $N_{i}$
impurity spins. However, as follows from Fig.~\ref{3imp}, only two of them 
are exact MZMs, namely those located at the right and left ends of the left and right
odd-number segments of the KM chain, respectively. The Majorana modes residing
at the boundaries of the inner (even-site) pieces of the KM chain overlap and split.
Nevertheless, in the dilute limit of rare impurities separated by distances
much larger than the correlation length,
the level splitting of the boundary modes
of each even-site inner segment is exponentially small, and
one can think of $2N_{i}$ MZMs in the disordered phase, forming $N_i$ nearly free spins. This is the limit
in which the interaction between the impurity spins can be neglected
and the symmetry of the disordered phase gets approximately promoted to 
$\left[ \mathbb{Z}_2 \right]^{N_i}$.
\medskip

\begin{figure}[!hbp]%[hbbp]
\centering
\includegraphics[width=2.4in]{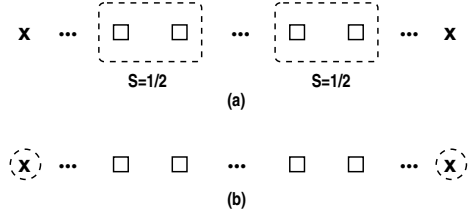}   %{many-imp.eps}
\caption{\footnotesize Majorana zero modes in a many-impurity system.
(a) disordered phase:  $2N_{\rm imp}$ Majorana quasi-zero modes;
(b) ordered phase: no localized free spins left, two exact MZMs at the end-points of the chain.  
}

\label{many-imp}
\end{figure}

%%%%%%%%%%%%%%%
%%%%%%%%%%%%%%%%
 The
difference between the numbers of the Majorana (quasi-)zero modes in the ordered and disordered 
phase of the dilute impurity system (see Fig.~\ref{many-imp}) originates from the
non-trivial topological property of the inner pieces of the KM lattice.
In spite of having
even numbers of $c$-sites, these pieces do \emph{not} represent parts of the original
spin chain defined in terms of the $\s^{\alpha}_n$ operators. Indeed, the sequence
of Majorana links in a finite QIC is $[hJhJ~ \cdots~ hJh]$, whereas in all inner even-site
pieces the sequence of links is different: $[JhJh~ \cdots~ JhJ]$. 
For these pieces it is impossible to form local spin operators by pairing the $c$-operators
according to the usual rule [12] [34] $\cdots$ [2M-1, 2M]. It can be readily seen that,
for the inner pieces of the KM lattice,
the pattern of pairing neighbors actually corresponds to the
\emph{dual} lattice of links, i.e., the lattice of the disorder $\mu^{\alpha}_n$-operators,
Eq.~(\ref{duality}).
Thus the inner even-site segments of
Fig.~\ref{3imp} can be treated as "physical" only in the dual ($\mu^{\alpha}_{j}$) representation.
In this sense, zero-field impurities favor a description using the
Kramers-Wannier duality transformation of the original quantum spin chain.

\medskip

As a consequence of this curious fact, in the inner regions
of the QIC separated by zero-field impurities, the zero boundary modes appear only in the disordered phase. Therefore for these regions it is the spin disordered phase
(being ordered in the dual representation) which is topological. This is 
consistent with the observation that for $J<h$ the boundary zero modes of neighboring
regions will combine and form local, free S=1/2 degrees of freedom.

%%%%%%%%%%%%%%

%%%%%%%%%%
\section{Conserved, free spin operator}\label{sec:conservedspin}
%%%%%%%%%%

As was  discussed in the preceding section, in %an 
a Quantum Ising model  the presence of a classical impurity with vanishing transverse field leads to an exact degeneracy of the entire spectrum, all eigenstates coming in pairs of equal energy, $|E_n,\pm\rangle$, where $\pm$ indicates the eigenvalue of $\sigma_0^x$. 
Formally one can thus define a set of three "spin operators" $S^{x,y,z}$ by their action in this basis: 
\bea
\label{freespinoperator}
S^z |E_n,\pm \rangle &=& |E_n,\mp\rangle,\\
S^y |E_n,\pm \rangle &=&  \mp i |E_n,\mp\rangle,\\ 
S^x|E_n,\pm \rangle &=& \pm |E_n,\pm\rangle.
\eea
They all commute with the Hamiltonian and satisfy the standard SU(2) commutation relations. These operators thus represent a free spin $1/2$ degree of freedom. Note however, that these latter properties do not uniquely determine the operators $S^{x,y,z}$. Indeed, we could have chosen the labelling of eigenstates such that $\pm$ refers instead to the eigenvalue of $\tau_n  \sigma_0^x$ with $\tau_n$ being a random sign $(\pm 1)$, chosen independently for each $n$. The same construction  then yields a different triplet of free spin operators. However, almost all of these choices would result in highly non-local operators, which have no practical interest. The above choice (\ref{freespinoperator}) 
with $\tau_n\equiv 1$  is singled out by the further requirement that the resulting operators be local in the ordered phase of the QIC. 
\medskip

It is not difficult to construct the operators $S^{x,y,z}$ explicitly using the Majorana zero modes on either side of the impurity. 
In terms of the notation introduced in Eq.~(\ref{not1-H}) {(see the next section)}, those zero modes  have the explicit expressions
\bea
\label{zero modes}
\Xi_a =  \sqrt{1-(J/h)^2}  \left( \beta_a +\sum_{j\geq 1}^L \eta_{aj} \left(\frac{J}{h}\right)^{j}\right), \quad a=1,2,
\eea
where for simplicity we consider the case of homogeneous Ising couplings, $\tilde{J}_{1,2}=J_{1,2}=J$.
The zero Majorana modes commute with the Hamiltonian (\ref{not1-H}), and for $h>J$ are normalized to satisfy $\Xi_{a}^2=1$ in the thermodynamic limit. In finite size systems the zero modes still exist even in the ordered phase, $h<J$, but there, the corresponding operators are dominated by the last terms in the sum (\ref{zero modes}) and thus act primarily on the far ends of the finite chains. 
\medskip

The bilinear 
\bea
S^z \equiv i \Xi_1 \Xi_2 
\eea
is obviously conserved, too, and squares to 1. Since in the product of  two Majorana operators the JW tails cancel, $S^z$ is a quasi-local operator in spin degrees of freedom for $h>J$, with the explicit representation
\bea
\label{explicitSz}
S^z = \left[1-\left(\frac{J}{h}\right)^2\right] \left[ \sigma_0^z -\sum_{j\leq 0\leq k; (j,k)\neq (0,0)} \left(\frac{J}{h}\right)^{k-j}\sigma_j^y \sigma_{k}^y \prod_{m=j+1}^{k-1} \sigma_{m}^z    \right].
\eea
Defining
\bea
S^x \equiv \sigma_0^x, 
\eea
one easily checks the anticommutation relation $\{S^z,S^x\}=0$. Defining eventually
\bea
S^y \equiv i S^x S^z,
\eea
one obtains a third conserved operator. It completes the triple of SU(2) operators which satisfies the spin algebra $\{S^i,S^j\}= 2 \delta _{ij}$ and $[S^i,S^j]= 2i \epsilon_{ijk} S^k$. The presence of this quasi-local free spin operator in the disordered phase implies not only the Curie form of the local susceptibility, but the two-fold degeneracy of the entire spectrum. Furthermore, 
to obtain the partner state of a given eigenstate which is simultaneously an eigenstate of $\sigma_0^x$, it suffices to act with $S^z$ or $S^y$ on that state. This establishes a specific local relationship between {\em all} many-body states, throughout the entire spectrum. Below we will use this property to estimate the local transverse susceptibility in the disordered phase, but close to criticality.

\subsection{The existence of conserved quasi-local spin operators is tied to integrability}
It is natural to ask whether the existence of a quasi-local, conserved $S=1/2$ spin operator is already implied by the mere presence of a classical impurity in the %a transverse field Ising 
QIC model. In any Ising model such an impurity always implies the two-fold degeneracy of the entire spectrum and, by the abstract construction above, the existence of some conserved operators $S^{x,y,z}$ which obey SU(2) commutation relations. However, we believe that these operators are generically non-local, except in models with strong disorder or with an integrable structure (as for the QIC). Indeed, consider the perturbative construction of the operator $S^z$ according to the following formal recipe. Consider the general Ising model,
\bea
H =  -\sum_{i} h_i \sigma_i^z -  \sum_{i,j} J_{ij} \sigma_i^x \sigma_{j}^x \equiv H_0 + H_1,
\eea
with the transverse field term being described by $H_0 \sim h$ and the exchange term by $H_1 \sim J \ll h$. The couplings $J_{ij}$ are non-zero only for  spins $i$ and $j$ that are spatially close, but not necessarily restricted to nearest neighbor pairs. Now make the perturbative ansatz
\bea
S^z \equiv S^{z,(0)} + S^{z,(1)} + \sum_{k\geq 2}S^{z,(k)} = \sigma_0^z - \sum_i \frac{J_{0i}}{h_i} \sigma_0^y \sigma_i^y  + O((J/h)^2) , 
\eea
where the norm of the operator $S^{z,(k)}$ scales as $(J/h)^k$, as $J\to 0$.
We may now try to find the $S^{z,(k)}$ iteratively by solving the conservation constraint $[H,S^z]=0$ order by order:
\bea
\label{PTk}
[H_0, S^{z,(k)}] + [H_1, S^{z,(k-1)}] = 0.    
\eea

This recipe has been followed in the context of many-body localized systems to construct quasi-local integrals of motion~\cite{Ros2015}. It has been argued that for sufficiently strong disorder (${\rm Var}({h_i})\gg J^2$, where $J$ is the typical nearest neighbor exchange coupling), the formal perturbation series defined by Eq.~(\ref{PTk}) can be resummed and leads to quasi-local integrals of motion. In this strongly disordered case, the construction actually works irrespectively of the value of $h_0$). The existence of such quasi-local conserved operators was  proven almost rigorously for one-dimensional Ising spin chains.~\cite{Imbrie}
\medskip

However, for this procedure to work, it is essential that sums and differences of sets of different 
$h_i$ do not vanish, and that they yield small values only with sufficiently low probability, since such terms appear in the denominators of the coefficients that multiply products of spin operators in $S^{z,(k)}$. In a homogeneous system, where all $h_i$ are equal (except for $h_0=0$), this requirement is maximally violated, and the above procedure is very likely to fail. In fact, one faces an extreme case of a small denominator problem, since many denominators of the formal perturbation theory will exactly vanish as $h_i\to h= {\rm const}.$.
This problem arises because the conjugation with $H_0$, $C(X) = [H_0,X]$  is not surjective as a linear map in operator space. 
For the case of generically disordered $h_i$ its kernel is however small enough, such that one can prove that a solution to (\ref{PTk}) can be found at every step.~\cite{Ros2015}
\medskip

However, for $h_i={\rm const}$, the kernel of $C$ is much larger, so that after a few steps of perturbation theory one cannot ensure that  $[H_1, S^{z,(k-1)}]$ lies in the image of $C$. This issue appears at the earliest in the 4$^{\rm th}$ order of perturbation theory, if triangles formed by exchange bonds are present. At this point, one would have to track back and seek for conserved operators using  degenerate perturbation theory, remembering that the exponentially large set of all polynomials in $\sigma_i^z$'s is conserved by $H_0$. They are thus all zero-eigenvectors under conjugation with $H_0$ and therefore, in general, perturbation theory has to be carried out starting from an appropriate linear combination of those degenerate eigen-operators. In view of the non-local nature of most of those polynomials, it seems very unlikely that such a procedure can be engineered to result in a local conserved operator $S^z$.
\medskip

Also from physical considerations there are good reasons  not to expect the existence of such a local conserved operator in general. A converging, quasi-local $S^z$ could be viewed as the creation operator of a sharp, quasi-local excitation with zero energy, that is, an exact zero-energy quasiparticle with infinite life time. In a non-integrable, non-localized system it is not conceivable that such infinitely long-lived excitations should exist. Indeed, at any finite temperature there is a finite phase space for the decay of that quasiparticle, upon scattering from excited delocalized modes  above the spectral gap. Of course, the cross-section of these processes decreases exponentially to zero as $T\to 0$. This ensures that ultimately, at $T=0$, there will still be a sharp excitation localized close to the impurity, and this will still give rise to a Curie susceptibility in the low $T$ limit. 
\medskip

However, for classical  impurities in very special, non-disordered Ising models such as a chain with only nearest neighbor interactions, the above-discussed problem of perturbation theory does not arise because of a lot of exact cancellations that kill the dangerous terms. A non-trivial example is given below.
The existence of such local operators in these systems is presumably tightly linked to the integrability of the transverse field Ising model in strictly 1d systems. Related issues in more complex, but integrable spin chains have recently been analyzed in Ref.~\onlinecite{Fendley2015}.
\medskip

Local conserved operators $S^z$ appear to exist only in fine-tuned systems, unless one considers strongly disordered, many-body localized systems. Nevertheless, having the explicit form of conserved $S^z$ in such fine-tuned systems helps us to visualize the physical properties associated with a classical impurity,  especially at low $T$. Furthermore, we expect those properties to be more general than the existence of the local $S^z$ operator itself. Since $S^z$ takes one ground state to another, the spatial extent of the operator makes precise the notion that the ground states differ only locally, a fact that should be generically true, independent of integrability or localization. 
Below, the structure of $S^z$ will also help us to understand the scaling of the low $T$ susceptibility upon approaching criticality.
The fact that in specific models $S^z$ as an operator is quasi-local implies additionally that {\em all} degenerate pairs of states are similar up to local modifications, not only the ground state.

\subsection{Exact conserved spin operators in junctions of transverse field Ising chains}

It is interesting to note that conserved free spin operators also exist in the case where the classical impurity sits at the junction of an arbitrary number $n$ of 1d chains - a situation in which integrability is not as obvious as in the cases of an impurity at the end or in the bulk of a simple chains, which correspond to $n=1,2$. Labelling the junction spin by $0$, one can check that the operator
\bea
\label{starSz}
S^z = \left(1-\frac{J^2}{h^2}\right)^{n/2}\sum_{k=0}^n (-i \sigma_0^x)^k \sigma_0^z \sum_{j_1,...,j_k\geq 1}\prod_{a=1}^k \left[\left(\frac{J}{h}\right)^{j_a}\sigma_{j_a}^y\prod_{i_a=1}^{j_a-1}  \sigma_{i_a}^z\right]
\eea
is indeed conserved by the Ising Hamiltonian on such a junction structure if $h_0=0$. This expression generalizes Eq.~(\ref{explicitSz}), to which it reduces for $n=2$. This operator also squares to $1$.
Again one obtains a full triplet of spin operators by completing $S^z$ with the conserved operators $S^x=\sigma_0^x$ and $S^y =i S^x S^z$.  
\medskip

In the low temperature limit, the Lehmann representation of the local spin susceptibility 
$\chi_0 = \lim_{h_0 \to 0} \left( \p \la \s^z _0 \ra \right)\p h_0$ leads to the Curie law
\be
\chi_0 (T) = \mu^2 _{\rm eff}/T, \label{CURIE}
\ee
where the effective magnetic moment $\mu_{\rm eff}$
is defined 
as the matrix element of $\sigma_0^z$ between the two degenerate ground states $|0,\pm\ra$,
\bea
\mu_{\rm eff} = \langle 0,-| \sigma_0^z | 0,+\rangle
= \la 0,+| S^z \sigma_0^z | 0,+\ra.
\label{mu-lehmann}
\eea
The states $|0,\pm\ra$  are eigenstates of $\sigma_0^x$ with eigenvalues $\pm 1$, respectively.
Inserting the explicit expression (\ref{starSz}) we see that the contribution from the leading term is simply 
\bea
\mu_{\rm eff} =  
\left(1-\frac{J^2}{h^2}\right)^{n/2}\Big|_{h-J \sim m\to 0} ~{\sim}~ m^{n/2}.
\eea
Assuming that the matrix elements of higher spin operator products come with random signs, one finds that they contribute with the same scaling to $\mu_{\rm eff}$. Close to criticality, in the continuum limit, one thus expects the scaling
\bea
\label{chi_n}
\chi_0(T\to 0) %= \frac{\mu_{\rm eff}^2}{T} 
\sim \frac{m^n}{T}.
\eea
This conjecture is confirmed by the explicit calculations for the cases $n=1$ and $n=2$, see Eqs.~(\ref{lowest-T}) and (\ref{m^2-scaling}).

%%%%%%%%%%%%%%%%

%%%%%%%%%%%%%%%%%%%%%%%%%%%%%%%%%%%%%
\section{Impurity in  a weakly non-critical QIC}\label{reduction-0}
%%%%%%%%%%%%%%%%%%%%%%%%%%%%%%%%%%%%%
\subsection{Reduction to a two-channel resonant-level model of massive Majorana fermions}\label{reduction}
%%%%%%%%%

In the rest of this paper we will be dealing with a single %classical-spin 
impurity
in a weakly non-critical QIC.
In this section we set up a formalism based on a continuum,  field-theoretical 
description of the bulk
degrees of freedom to treat effects caused by the impurity spin. 
The impurity is located at the origin. The right and left
parts of the chain, supplied with subscripts 1 and 2, respectively, are assumed to be homogeneous, 
but may represent 
different quantum Ising chains characterized by two sets of
parameters, ($J_1, h_1$) and ($J_2, h_2$). In the KM representation, the Hamiltonian of the system reads:

%%%%%%%%%%%
\bea
H &=& - \ri h_0 \beta_1 \beta_2 + \ri \tilde{J}_1 \beta_1 c_1 + \ri \tilde{J}_2 c_{-2} \beta_2 \nn\\
&+& \ri J_1 ( c_2 c_3 + c_4 c_5 + \cdots )
+ \ri h_1 ( c_1 c_2 + c_3 c_4 + \cdots )\nn\\
&+& \ri J_2 ( c_{-4}c_{-3} + c_{-6}c_{-5} + \cdots )
+ \ri h_2 ( c_{-3} c_{-2} + c_{-5} c_{-4}
+ \cdots )
\label{notA-ham}
\eea
The pair of Majorana operators,
$\beta_1$ and $\beta_2$ (previously denoted $c_0$ and $c_{-1}$, respectively)  describes the impurity spin $\s^z _0 = \ri \beta_1 \beta_2$
located at the origin.
$h_0$ is the %local 
transverse magnetic field at the impurity site
which must be kept small but finite to calculate the local response function.
The coupling constants  $\tilde{J}_1$ and $\tilde{J}_2$ parametrize hybridization between
the impurity Majorana fermions and the bulk degrees of freedom.

%%%%%%%%%%%%%
\medskip

\begin{figure}[hbbp]
\centering
\includegraphics[width=2.5in]{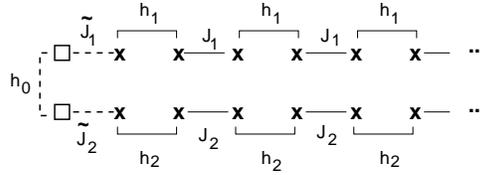}
\caption{\footnotesize Folded KM chain with an impurity.}
\label{folded}
\end{figure}

Before passing to the continuum limit, it is convenient to distinguish between even and odd sites of the KM lattice. 
For the right part of the chain we set
$c_{2j-1} = \zeta_j$, $c_{2j} = \eta_j$ ~$(1 \leq j \leq N)$
with a minor modification for the left part, 
$c_{2j-1} = \eta_j$, $c_{2j} = -\zeta_j$  $( -N \leq j \leq -1)$. 
In the new notations the Hamiltonian reads %(see Fig.~\ref{maj-lat-imp})
\bea
H &=& - \ri h_0 \beta_1 \beta_2 + \ri \tilde{J}_1 \beta_1 \zeta_1
+ \ri \tilde{J}_2 \beta_2 \zeta_{-1} 
+ \ri \sum_{j \geq 1}\left( h_1\zeta_i \eta_j + J_1 \eta_j \zeta_{j+1} \right)
- \ri \sum_{j \geq 1}\left( h_2\eta_{-j} \zeta_{-j} + J_2  \zeta_{-j-1} \eta_{-j}\right).
\nn 
\eea
Mapping the negative semi-axis to the positive one, as shown
in Fig.~\ref{folded}, and introducing two
species of Majorana operators, %using the identification
$
\eta_j \equiv \eta_{1j}, ~\zeta_j \equiv \zeta_{1j}, ~
\eta_{-j} \equiv \eta_{2j}, ~\zeta_{-j} \equiv \zeta_{2j},
$
we transform the Hamiltonian as follows 
\bea
H = - \ri h_0 \beta_1 \beta_2 + \ri \sum_{a=1,2}\tilde{J}_a \beta_a \zeta_{a1}
+  \ri \sum_{a=1,2} \sum_{j \geq 1}\left(J_a \eta_{aj} \zeta_{a,j+1}  -
h_a \eta_{aj} \zeta_{aj}  \right).\label{not1-H}
\eea
\medskip

\noindent
The last term in
(\ref{not1-H}) represents a sum of two semi-infinite QIC models.
Assuming that each chain is close to criticality,
$|h_a-J_a|\ll J_a$$~(a=1,2)$,
one can pass from the lattice Majorana operators to continuum fields using the correspondence
\be
\eta_{aj} \to \sqrt{2a_0}~ \eta_a (x), ~~~
\zeta_{aj} \to \sqrt{2a_0}~ \zeta_a (x). \label{eta-zeta-cont}
\ee
The fields satisfy the algebra
\bea
\{ \eta_a (x), \eta_b (x') \} = \{ \zeta_a (x), \zeta_b (x') \} =
\delta_{ab} \delta(x-x'), ~~~
\{ \eta_a (x), \zeta_b (x') \} = 0.
\label{eta-zeta-a-alg}
\eea
In the continuum limit, the Hamiltonian (\ref{not1-H}) takes the form
\bea
H = - \ri h_0 \beta_1 \beta_2  ~+~ \ri \sqrt{2a_0} \sum_{a=1,2}\tilde{J}_a  \beta_a \zeta_a (0)
+\sum_{a=1,2} \int_0 ^L \rd x~\left[\ri v_a \eta_a (x) \p_x \zeta_a (x)
- \ri m_a \eta_a (x) \zeta_a (x)  \right].~~~~\label{H-fin-rep}
\eea
In (\ref{H-fin-rep}) the bulk degrees of freedom of the system are described in terms
of two (formally Lorentz-invariant) free massive Majorana fields with group velocities
$v_a = 2J_a a_0$ and masses $m_a = 2 (h_a - J_a)$.
\medskip

In the continuum limit, the Majorana fields at the open end must satisfy boundary conditions which are obtained as follows.
Take a semi-infinite KM lattice $(\zeta_1, \eta_1), (\zeta_2, \eta_2), \ldots$
and add an extra site $j=0$ requiring that %there should be no 
the pairing between $\eta_0$ and $\zeta_1$ is absent. For the degrees of freedom on the chains this is equivalent to having a finite coupling, but imposing the constraint $\eta_0 =0$.
Thus, in the continuum description, which captures the relevant low-energy
subspace of the model, the boundary conditions are
\be
\eta_a (0) = 0, ~~~(a=1,2).
\label{bc-s}
\ee
The original model is thus equivalently represented as
a Majorana version of the two-channel, massive resonant-level model on a semi-infinite (for $N \to \infty$) axis with
an impurity at the boundary. It is crucial that 
different bulk channels ($a=1,2$) are coupled
to different Majorana components of the impurity spin ($\beta_1$ and $\beta_2$).
Note that if $h_0 = 0$ the 
channels decouple.
\medskip

The massless limit of the model (\ref{H-fin-rep}) has 
been thoroughly studied
long ago in the context of the overscreened  Kondo effect. Indeed, at
$m_1 = m_2 = 0$ and $v_1 = v_2$,
Eq.~(\ref{H-fin-rep}) 
represents the %two-channel 
Majorana resonant-level model introduced and
solved by Emery and Kivelson~\cite{EK} in their bosonization treatment of
the two-channel Kondo problem with XXZ exchange anisotropy 
(see e.g. Ref.~\onlinecite{book} for a review). 
The correspondence between the critical point of the model (\ref{H-fin-rep}) 
and the two-channel resonant-level model %Kondo problem
is briefly discussed in Appendix \ref{crit}.
When one of the two bulk-impurity couplings, $\tilde{J}_1$ or $\tilde{J}_2$, vanishes,
one has a %single-channel 
problem of massive Majorana fermions on a single semi-infinite chain with 
an impurity at the open end. In the related
two-channel Kondo model, this case corresponds to 
the channel-symmetric situation describing the overscreened Kondo effect.
For two identical massless chains %($m_1 = m_2 = 0$, $v_1 = v_2$) %($J_1 = J_2$, $h_1 = h_2$) 
and coinciding
coupling constants ($\tilde{J}_1 = \tilde{J}_2$) the Hamiltonian (\ref{H-fin-rep})
acquires an enhanced, $O(2)$, symmetry
associated with global planar rotations of the first and second Majorana species.
In this case the Hamiltonian (\ref{H-fin-rep}) %actually 
represents the usual
resonant-level model for complex fermions, in which the total particle number is 
conserved and which is relevant to the single-channel Kondo problem\cite{and}. 

At $m_1 = m_2 \neq 0$, $v_1 = v_2$ the model (\ref{H-fin-rep}) describes a semi-infinite 
 Peierls insulator (PI) cahin with an impurity fermionic $d$-state at the open boundary whose
energy is $\vare_d = - 2h_0$ (see Appendix \ref{peierls}).
\medskip

%%%%%%%%%%%%%%%%%%%%%%%%%%%%%%%%%%%%%%%%%%

\subsection{Diagonalized bulk spectrum and the total Hamiltonian}

To diagonalize the bulk part of the Hamiltonian (\ref{H-fin-rep}), one first performs
a chiral rotation of the Majorana fields
$(\eta_a, \zeta_a )$:
\bea
\eta_a (x) = \frac{\xi_{aR} (x) + \xi_{aL} (x)}{\sqrt{2}}, ~~~~~
\zeta_a (x) = \frac{- \xi_{aR} (x) + \xi_{aL} (x)}{\sqrt{2}}.
\label{xi-s}
\eea
The new Majorana fields satisfy the algebra:
\bea
\{ \xi_{aR} (x), \xi_{bR} (x') \} = \{ \xi_{aL} (x), \xi_{bL} (x') \}
= \delta_{ab} \delta(x-x'),~~~~
\{ \xi_{aR} (x), \xi_{bL} (x') \} = 0. 
\label{xi-algebra}
\eea
The boundary conditions (\ref{bc-s}) translate to
\be
\xi_{aR} (0) = - \xi_{aL} (0). \label{bc-xi}
\ee
In terms of the  fields $\xi_{a;R,L} (x)$ the Hamiltonian takes the form
\bea
&&H  - \ri h_0 \beta_1 \beta_2  - 2\ri \sqrt{a_0} \sum_{a=1,2}\tilde{J}_a  \beta_a 
\xi_{aR} (0) ~~~~\nn\\
&& +\sum_{a=1,2} \int_0 ^L \rd x~\Big[\frac{\ri v_a}{2} \left(\xi_{aL} (x) \p_x \xi_{aL} (x)
- \xi_{aR} (x) \p_x \xi_{aR} (x)\right)
- \ri m_a \xi_{aR} (x) \xi_{aL} (x)  \Big].
\label{H-fin-rep-xi}
\eea

The diagonalization of a massive Majorana model on a semi-axis is discussed in Appendix 
\ref{diagonalization}. The spectrum contains a continuum of extended states
with the energy $\vare_{k} = \sqrt{k^2 v^2 + m^2}$ and, for a negative mass only
($m<0$), a normalizable zero-energy state localized at the boundary of the chain
within a characteristic length $\xi = a_0/|\ln(h/J)| \simeq a_0/(1-h/J) = v/|m|$. The normal-mode expansion of
a single massive Majorana spinor field operator is given by formula (\ref{xi-expan3}).
Using this expansion one rewrites the Hamiltonian as follows:
\bea
H = - \ri h_0 \beta_1 \beta_2 + \sum_{k>0}\sum_a  \vare_{ak} \hat{\gamma}^\dagger _{ak} 
\hat{\gamma}_{ak} 
- \ri \sum_a \lambda_0^a \beta_a \gamma_{a0} 
- \ri \sqrt{\frac{2}{N}} \sum_a \beta_a \sum_{k>0} \lambda^a _k (\gamma_{ak} 
+ \gamma_{ak}^{\dagger} ).~~~\label{total-ham}
\eea
Here $\gamma_k, ~\gamma^{\dagger}_k$ are fermionic quasiparticle operators 
related to the continuum part of the spectrum, $\gamma_0 = \gamma^{\dagger}_0$ is the
operator describing the boundary MZM, and
\bea
\lambda_0^a = 2 \theta(-m_a)\sqrt{\frac{|m_a|a_0}{v}} \tilde{J}_a, ~~
\lambda^a _k = \tilde{J}_a \frac{kv}{\vare_{ak}}
\label{lambdas}
\eea
are the coupling constants.
In the last term of (\ref{total-ham}) the local Majorana operators $\beta_a$
couple to a Hermitian combination of the band operators, $\gamma_{ak} 
+ \gamma_{ak}^{\dagger} $. Therefore we can "majoranize" the bulk part of the spectrum as well.
Represent the quasiparticle operator $\gamma^{\dagger}_k$ as a linear combination
of two
real operators:
$
\gamma^{\dagger}_k = (b_k + \ri f_k)/2.
$
%where $b$ and $f$ are real operators. 
The algebra
$\{ \gamma_k, \gamma^{\dagger}_{k'} \} = \delta_{kk'}$ implies that
$
\{ b_k, b_{k'} \} = \{ f_k, f_{k'} \} = 2\delta_{kk'}
$
with all remaining anticommutators vanishing. %The new Majorana operators also satisfy
The kinetic energy becomes
\bea
\sum_{k>0} \vare_k \gamma^{\dagger}_{k} \gamma_k
= - \frac{\ri}{2} \sum_{k>0} \vare_k b_k f_k + {\rm const}.
\nn
\eea
Thus the total Hamiltonian can be represented entirely in terms of Majorana degrees if freedom:
\bea
 H_M = - \ri h_0 \beta_1 \beta_2 - \frac{\ri}{2} \sum_{k>0, a} \vare_{ak} b_{ak} f_{ak}
- \ri \sum_a \lambda_0^a \beta_a \gamma_{a0} 
- \ri \sqrt{\frac{2}{N}}\sum_a \sum_{k>0} \lambda^a _k \beta_a b_{ak}.~~~~
\label{maj-ham-final}
\eea
As we shall show below, the coupling of the impurity Majorana operators
$\beta_a$ to the Majorana boundary MZM operators $\gamma_{a0}$, %representing boundary zero modes of
parametrized by the constants $\lambda_0 ^a$, plays
a crucial role in the low-temperature asymptotics of the local spin susceptibility
in the ordered bulk phase.
%%%%%%%%%%%%%%%%%%%%%%%%%%%%%%%%%%%%

%%%%%%%%%%%%%%%%%%%%%%%%%%%%%%%%%%%%

\subsection{Physical quantities in terms of Green's functions}

The Hamiltonian $H_M$ in (\ref{maj-ham-final}) represents a one-particle, exactly solvable
model. Our goal is to calculate the impurity parts of the physical quantities: 
the spectral weight 
of the impurity spin, the local magnetization  and spin
susceptibility defined as the linear response to the local transverse
magnetic field.
All calculations are straightforward
and can be done using the formalism of Matsubara Green's functions (GF)\cite{AGD}. 
\medskip

Here we provide basic definitions. We remind that
the impurity spin is described in terms of two Majorana operators $\beta_1$ and $\beta_2$:
$\s^x _0 = \beta_2$, ~$\s^y _0 = \beta_1$, ~$\s^z _0 = \ri \beta_1  \beta_2$, % \label{sigma-beta}\\
or equivalently, a complex (Jordan-Wigner) spinless $d$-fermion:
\bea
\label{spin-fermion-imp}
\s^z _0 = 2d^{\dagger} d - 1 ~\equiv~ 2n_d-1, ~~~~\s^+ _0 &=& 2d^{\dagger} \exp[i \pi N_2], ~~~~\s^- _0 = 2d \exp [i \pi N_2],\\
  N_2 &=& \sum_{j=1}^{\infty} a^{\dagger}_{2,j} a_{2,j}.
\eea
A local magnetic field $h_0$ determines the Zeeman energy
of the impurity spin,
$
H_0 = - h_0 \s^z _0.
$
Accordingly, the local magnetization and spin susceptibility are defined as
\bea
 m_0 (h_0, T) = \la  \s^z _0 \ra = 2 (\la n_d\ra  - 1/2), ~~~~
\chi_0 (T) = \frac{\p m_0}{\p h_0}\Big|_{h_0 = 0}\,.
\label{m-chi-def}
\eea
In the context of a 1DPS, the local magnetization of the impurity in a QIC translates 
to the average occupation number 
of the $d$-fermionic state,
$m_0 \to 2 (\la n_d \ra- 1/2)$, the magnetic field $h_0$ transforms to the local
energy of the $d$-fermion, $\mu_0 = 2h_0$, and the local spin susceptibility $\chi_0$
becomes the local charge susceptibility ("compressibility") of the impurity site. 
\medskip

An  important characteristics of free fermionic models
which determines local thermodynamic properties of the impurity
is  the spectral weight (or local density of states) of the boundary complex  fermion:
\be
A(\omega) = - \frac{1}{\pi} \Im m~{\cal G} (\omega + \ri \delta), ~~~~~~
\int_{-\infty}^{\infty} \rd \omega~ A(\omega) = 1. \label{A-via-G}
\ee
Here the retarded GF ${\cal G} (\omega + \ri \delta)$ is the analytic continuation of the
Fourier transform $G(\vare_n)$ of the Matsubara single-fermion GF %\cite{AGD}
\bea
G(\vare_n) = \int_0 ^{1/T} \rd \tau~e^{i\vare_n \tau} G(\tau),~~~
G(\tau) = - \la  T_{\tau} d(\tau) d^{\dagger} \ra, ~~~~\vare_n = (2n+1) \pi T
\label{G}
\eea
where $T_{\tau}$ is the imaginary-time ordering operator and
$d(\tau) = e^{\tau H} d e^{-\tau H}$. 
Using the integral representation for $G(\vare_n)$
\be
G(\vare_n) = \int_{-\infty}^{\infty} \rd \omega~\frac{A(\omega)}{\ri \vare_n - \omega},
\label{G-via-A-int}
\ee
one obtains the expressions for $m_0$ and $\la n_d \ra$\cite{AGD}
\bea
m_0 (h_0,T) =- \int_{-\infty}^{\infty} \rd \omega ~A(\omega; h_0) \tanh\frac{\omega}{2T},~~~~~
%\label{m-via-A}\\
\la  n_d \ra = \int_{-\infty}^{\infty} \rd \omega ~A(\omega; h_0) f(\omega)
\label{m-n-via-A}
\eea
where $f(\omega) = (e^{\omega/T} +1)^{-1}$ is the Fermi distribution function.
\medskip

{
Thus, the spectral weight determines the local magnetization $m_0$ and average occupation 
$\la n_d \ra$ of the impurity site, both being experimentally accessible quantities. However,
when describing the impurity spin dynamics in terms of the local
$d$-fermion, one should pay attention to the important difference between the 
cases of a boundary impurity in a semi-infinite QIC  and an impurity in the bulk
of the spin chain.
In the former case, the impurity spin is located at the open end of the chain, so
%(i.e. at the 0-site); there 
the boundary spin operators $\s^{\pm}_0$ do not contain the
JW exponentials % tails 
and are \emph{locally} expressed in terms of $d$ and $d^{\dagger}$:
$\s^+ _0 = 2d^{\dagger}$, $\s^- _0 = 2d$.
In this case the $d$-fermion Green's function $G_d(\tau)$ coincides with the spin-spin correlation 
function,
\be
G_d (\tau) = - \frac{1}{4} \la \s^- _0 (\tau) \s^+ _0 (0) \ra,
\label{Gd1}
\ee
and the spectral function $A(\omega)$
measures the fluctuation spectrum
of the impurity spin.}
\medskip

{
This is not so for an impurity coupled to both Ising chains, $\tilde{J}_1, \tilde{J}_2 \neq 0$.
In this case, only the total fermion number parity
$P_S = \exp [i \pi (N_1 + N_2)]$ is conserved , while
the parities of each chain, $P_{1,2}$, are not. Therefore, in the definition (\ref{spin-fermion-imp}),
the JW "tail" operator $P_2 = \exp (i\pi N_2)$ has a nontrivial dynamics, implying that
for an impurity in the bulk of the QIC, the impurity spin
components $\s^{\pm}_0$ are essentially nonlocal objects in terms of the JW fermions.  
In this case $A(\omega)$ cannot be expressed in terms of simple local spin-spin correlators
and thus only has the meaning of a density of states for the $d$-fermion.}
\medskip

\medskip

At $h_0 =0$ the Hamiltonian (\ref{maj-ham-final}) has an exact sub-chain %particle-hole 
symmetry:
it remains invariant under the transformations 
\bea
&& \beta_1 \to - \beta_1,~~\gamma_{10} \to  {-} \gamma_{10}, ~~b_{1k} \to - b_{1k}, ~~f_{1k} \to - f_{1k},~~~
\nn\\
&& \beta_2 \to \beta_2,~~\gamma_{20} \to \gamma_{20}, ~~b_{2k} \to  b_{2k}, ~~f_{2k} \to f_{2k}.
~~\label{ph}
\eea
In particular,  the transformation (\ref{ph}) swaps $d \leftrightarrow d^{\dagger}$, implying that at $h_0 = 0$
$G(\vare_n) = - G(-\vare_n)$ and, according to (\ref{G-via-A-int}), $A(\omega;0) = A(-\omega;0)$.
Therefore at any finite temperature $m_0 \to 0$ as $h_0 \to 0$.
However, the limit $h_0 \to 0$ does in general not commute with the limit $T\to 0$, in which $\tanh (\omega/2T)$ acquires a discontinuity
at $\omega = 0$. The result of the integration in (\ref{m-n-via-A})
thus becomes ambiguous if $A(\omega)$ has a
$\delta$-function singularity at $\omega =0$. 
This singularity reflects the two-fold degeneracy
of the impurity ground state. The local zero-temperature
magnetization will exhibit a discontinuity as $h_0$, and with it the $\delta$-function singularity of $A(\omega)$, cross zero.
A free spin 1/2 is a simple example of this kind: when the external
magnetic field is switched off, the 
magnetization $m_0 (h_0) = {\rm sgn}(h_0)$ keeps track of its original orientation.
Switching on a finite magnetic field $h_0$
generates an antisymmetric part of the spectral weight which yields a nonzero contribution
to the integral (\ref{m-n-via-A}).
The spin susceptibility can then be obtained using the definition (\ref{m-chi-def}).
Formula (\ref{m-n-via-A}) for $m_0$ will be  used extensively below. 
\medskip

The zero-field local susceptibility $\chi_0(T) = [\p m_0/ \p h_0]_{h_0 = 0}$
has an equivalent representation in terms of the response function
\be
\chi^{zz} (\omega) = \ri \int_0 ^{\infty} \rd t~e^{i\omega t}
\la  [\s^z _0 (t),  \s^z _0 (0)]\ra . \label{resp-funct}
\ee
Its Matsubara counterpart 
\be
X^{zz}_0 (\omega_m) = \int_0 ^{1/T} \rd \tau~e^{i\omega_m \tau}
\la  T_{\tau} \s^z _0 (\tau)  \s^z _0 (0)\ra,
 ~~~~(\omega_m = 2m\pi T), \nn%\label{X_m}
\ee
represents a "polarization loop" of two impurity Majorana GFs:
\bea
X^{zz} _0 (\omega_m) = T \sum_{\vare_n} \left[ D_{11}(\vare_n) D_{22} (\omega_m - \vare_n)
- D_{12}(\vare_n) D_{21} (\omega_m - \vare_n)\right]. \nn
%\label{X-loop}
\eea
{$\chi_0 (T)$ is defined as the static limit of the local dynamical spin susceptibility:
\be
\chi_0(T) = X^{zz}_0 (\omega_m = 0) = T \sum_{\vare_n} \left[
D_{11}(\vare_n) D_{22} (- \vare_n) - D_{12}(\vare_n) D_{21} (-\vare_n) \right].
\label{chi-loop-gen}
\ee
In the model (\ref{maj-ham-final}), at $h_0 = 0$ the Majorana fields $\beta_1$ and $\beta_2$
are decoupled, and in (\ref{chi-loop-gen}) one should set $D_{12} = D_{21} = 0$.
The representation (\ref{chi-loop-gen}) will prove useful in the next section when we analyze the susceptibility close to criticality.}

%%%%%%%%%%%%%%%%%%%%%

%%%%%%%%%%%%%%%%%%%%%%%%%%%%%%%%%%%%%%%%%%%%%%%%%%%%%%%%%%%%%%%%%
\section{Boundary impurity in a semi-infinite quantum Ising chain}\label{semi}

In the remainder of this paper we  primarily deal with a situation
displaying rich physics at the boundary:
the model of a single non-critical semi-infinite QIC with an impurity spin
at the open end. It is obtained from the Hamiltonian
(\ref{notA-ham}) by cutting off the coupling of the impurity Majorana fermion
$\beta_2$ to the second channel ($\tilde{J}_2 = 0$). 
We will first gain some intuition about the MZMs in the two phases in the discrete version of this model
and then turn to a continuum desciption assuming that massive phases of the semi-infinite QIC
are only weakly non-critical. 
\medskip

A qualitative picture of the MZMs can be inferred 
from Fig.~\ref{folded}
where  the lower KM chain should be completely ignored.
Set $h_0 = 0$, $\tilde{J}_1 \neq 0$
and consider the limiting cases $J/h \to 0$ and $h/J \to 0$.
In the former case, the impurity
degrees of freedom are represented by two MZMs: one being the decoupled Majorana fermion $\beta_2$ and
the other contained in the spectrum of the 3-site complex, $H_3 = \ri \tilde{J}_1 \beta_1c_1 
+ \ri h c_1 c_2$.
Therefore in the disordered phase ($J > h \neq 0$) there are two zero-energy
bound states
attached to the boundary: one of them  localized
just at the impurity site %the end of the chain 
while the other one has a finite penetration into
the bulk. The two MZMs describe a well-defined local spin degree of freedom.
\medskip

In the ordered phase, there remains only one MZM ($\beta_2$)
localized exactly at the boundary. The second impurity Majorana
($\beta_1$) couples with a nonzero binding energy to the bulk fermion $c_1$, which, in the impurity-bulk
decoupling limit 
($\tilde{J}_1 = 0$), represents
the boundary MZM of an isolated QIC in its ordered phase~\cite{kitaev2}. %by $\tilde{J}$. 
In this phase no local spin 1/2 remains at the impurity site.
As we will see in what follows, the bound state between the impurity Majorana fermion and the boundary zero mode of
the bulk %part of the spectrum 
plays a crucial role in  
quenching the transverse susceptibility
of the impurity
spin in the topological (ordered) phase of the QIC.
\medskip

%%%%%%%%%%%%%%%%%%%%%%%%%%%%%%%%

\subsection{Spectral weight of the impurity spin}
\label{spec-w}
%%%%%%%%%%%%%%%%%%%%%%%%%
Now we turn to a continuum model describing a semi-infinite, weakly non-critical
QIC with an impurity spin at the open end. Such a model
is obtained from (\ref{maj-ham-final}) by 
removing
the coupling of the %impurity 
Majorana fermion
$\beta_2$ to the second channel ($\lambda_{20} = \lambda _{2k} = 0$): 
\bea
H = - \ri h_0 \beta_1 \beta_2 - \frac{\ri}{2} \sum_{k>0} \vare_{k} b_{k} f_{k} 
- \ri  \lambda_0 \beta_1 \gamma_{0} 
- \ri \sqrt{\frac{2}{N}}\sum_{k>0} \lambda _k \beta_1 b_{k}.
\label{maj-ham-final12}
\eea
Here $\lambda_0 = \lambda_{10}$, $\lambda_{k} = \lambda_{1k}$, $b_k = b_{1k}$,
$f_k = f_{1k}$. The GF $G(\vare_n)$ is calculated in Appendix \ref{derivGF}:
\bea
 G(\vare_n, h_0) = \frac{1}{2i \vare_n} \left[ 1 - \frac{(\ri \vare_n - 2h_0)^2}
{\Delta(\vare_n) + 4h^2 _0} \right].
\eea
Here
\[
\Delta(\vare_n) = \vare^2 _n + \Gamma \left(  \sqrt{\vare^2 _n + m^2} - m \right),
\]
where $\Gamma = 4 \tilde{J}^2 a_0 / v = 2\tilde{J}^2/J$ is the hybridization width of the impurity
level.
\medskip

Analytic continuation $G(\vare_n)\to {\cal G} (\omega + \ri \delta)$
should be done according to the prescription: 
under $\ri \vare_n \to \omega + \ri \delta ~(\vare_n > 0)$
\bea
\sqrt{\vare^2 _n + m^2} &\to& - \ri \sqrt{\omega^2 - m^2} \,{\rm sgn}~\omega, 
~~~~{\rm if}~\omega^2 > m^2, \nn\\
&\to& ~\sqrt{m^2 - \omega^2}, ~~~~~~~~~~~~~~{\rm if}~~\omega^2 < m^2.
\label{anal-cont}
\eea
This yields %Using this prescription, one derives 
the following expressions for
the retarded GF %which has a different structure
at high and low frequencies:
\bea
&&\omega^2 > m^2: ~{\cal G}_>(\omega + i \delta; h_0) 
= \frac{1}{2(\omega + \ri \delta)}
\left[
1 + \frac{(\omega - 2h_0)^2}{\omega^2 + \Gamma m - 4h^2 _0 + \ri \Gamma \sqrt{\omega^2 - m^2}
~{\rm sgn}~\omega}
\right],\label{cal-G>}
\\
\nn\\
&&\omega^2 < m^2: ~{\cal G}_<(\omega + i \delta; h_0) = \frac{1}{2(\omega + \ri \delta)}
\left[
1 + \frac{(\omega - 2h_0)^2}{\omega^2 + \Gamma m - 4h^2 _0 - \Gamma \sqrt{m^2 - \omega^2}
+ \ri \delta~ {\rm sgn}~\omega} \right]. ~~~
\label{cal-G<}
\eea
Separating the imaginary parts of the GFs  (\ref{cal-G>}) and (\ref{cal-G<})
we  find that, at $h_0 = 0$, for a non-critical semi-infinite QIC the impurity spectral weight
has the following form:
\be
A(\omega) = \frac{1}{2} \delta (\omega) +
\frac{1}{2} A_{\rm f} (\omega).
\label{A-relation}
\ee
Here the $\delta$-function term is the contribution of the impurity Majorana fermion $\beta_2$,
which is completely decoupled at $h_0 = 0$. On the other hand, $A_{\rm f} (\omega)$ alone
represents the impurity spectral weight for two identical QICs symmetrically
coupled to the impurity, in which case the model maps 
onto the standard massive 
semi-infinite resonant-level model. 
In particular, such a model emerges in a continuum description
of a 1D spinless Peierls insulator %(PI) 
which also possesses topological and trivial massive phases\cite{shen}
(see Appendix \ref{peierls}).
The
explicit expressions for 
the impurity spectral weight $A_{\rm f}(\omega)$ at $h_0 = 0$ are:
\bea
&& m>0:~~
A_{\rm f}(\omega) ~=~ Z \delta (\omega)
~+~ \theta ~(\omega^2 - m^2) \frac{\Gamma}{\pi}
\frac{\sqrt{\omega^2 - m^2}}{|\omega| \left( \omega^2 +2\Gamma m + \Gamma^2 \right)};
\label{A-m>0}
\\
&&m<0: ~~
A_{\rm f} (\omega) ~=~  \theta (|m| - \Gamma) \frac{1}{2}\left( \frac{|m|-\Gamma}{2|m|-\Gamma} \right)
\left[
\delta(\omega - \omega_0) + \delta(\omega + \omega_0)\right]\nn\\
&& ~~~~~~~~~~~~~~~~~~~~~+~ \theta (\omega^2 - m^2) \frac{\Gamma}{\pi}
\frac{\sqrt{\omega^2 - m^2}}{|\omega| \left( \omega^2 - 2\Gamma |m| + \Gamma^2 \right)};\label{A-m<0}
\eea
where
\bea
Z &=& \frac{2m}{\Gamma  + 2m},~~~~~~~~~~~~~~~~~~~~~(m>0),\label{Z}\\
\omega_0 &=& \sqrt{\Gamma (2|m| - \Gamma)} < |m|,~~~~~(m < 0 ~~ {\rm and}~~ |m|>\Gamma). \label{o0}
\eea
Below we split the susceptibility into contributions from the discrete $\delta$-functions, $\chi_<$, and from the continuum ($\omega^2> m^2$) above the gap, $\chi_>$, respectively, 
\be
\chi_0 = \chi_< + \chi_>.
\ee

For a critical Majorana resonant level model (the case $m=0$)
the additive structure of the spectral weight $A(\omega)$ given by the sum (\ref{A-relation})
was first obtained by Emery and Kivelson \cite{EK}. As we see, the immunity of the decoupled
Majorana fermion $\beta_2$ keeps the structure (\ref{A-relation}) valid in the massive
case as well. It shows that only one half of the
boundary spin degrees of freedom is hybridized with the gapless bulk excitations while the
other half is decoupled from the rest of the system. 
\medskip

The representation (\ref{A-relation}) is no longer valid at $h_0 \neq 0$.
The spectral weight at a small nonzero $h_0$ will be discussed separately in the sequel.

\subsection{Local magnetization and spin susceptibility}\label{mag-suscep} 
\subsubsection{Critical state, $m=0$}

At $m=0$ the Hamiltonian (\ref{maj-ham-final12}) coincides with the resonant-level model
introduced and solved by Emery and Kivelson \cite{EK} in their treatment of the two-channel Kondo
problem. For the sake of completeness and later comparison with massive cases,
we reproduce their main findings here.
\medskip

The spectral weight  of the $d$-fermion is given by formula (\ref{A-relation}).
Setting $m=0$ in Eq. (\ref{A-m>0}) or (\ref{A-m<0}) one obtains\cite{EK}
($h_0 = 0)$
\be
A(\omega) = 
\frac{1}{2} \delta(\omega) + \frac{1}{2} \frac{\Gamma}{\pi} \frac{1}{\omega^2 +\Gamma^2}.
\label{A-EK}
\ee
The zero-temperature entropy of the impurity
spin, $S = (1/2)\ln 2$, is a half of its value for an isolated spin 1/2.
This is a direct effect of the decoupled
boundary  Majorana $\beta_2$-fermion. Applying a small local magnetic field $h_0$
induces %the appearance of 
an antisymmetric part of $A(\omega)$ 
which can be easily obtained from (\ref{cal-G>}).
According to the
definition (\ref{m-n-via-A}), this leads to the following result for the zero-field local susceptibility
\bea
\chi_0 (T) = \frac{4\Gamma}{\pi}\int_0 ^{\infty}\frac{\rd \omega}{\omega}
\frac{\tanh(\omega/2T)}{\omega^2 + \Gamma^2}.
\label{chi-crit-integral}
\eea
At $T \gg \Gamma$
the susceptibility follows the Curie law,
$
\chi_0 =  1/T,
$
as expected. In contrast,
at low temperatures, $T \ll \Gamma$, it has a logarithmic temperature
dependence typical for the two-channel, overscreened Kondo impurity in the channel-symmetric 
case~\cite{EK}:
\bea
\chi_0 (T) = \frac{4}{\pi \Gamma}\ln \frac{\Gamma}{T}.
\label{2H-log}
\eea

It is instructive 
to make a comparison with the case of two (instead of one) identical critical Majorana chains
coupled to the boundary spin. This
reduces to the
standard resonant-level model describing a semi-infinite metallic chain 
with a $d$-fermion state at its boundary.
Such a model with two attached critical Ising chains maps identically to
a U(1)-symmetric spin-1/2, semi-infinite XX chain
with a boundary spin 1/2 in a transverse magnetic field $h_0$.
This model is known to be related to the standard, one-channel Kondo problem at the
Toulouse point (see e.g. Ref.~\onlinecite{book}).
In this case the impurity spectral weight and the local charge susceptibility at
$T/\Gamma \to 0$, %are given by
\bea
A_{\rm f}(\omega) = \frac{\Gamma}{\pi} \frac{1}{\omega^2 + \Gamma^2}, ~~~~
\chi = \frac{1}{\pi \Gamma}, \label{A-chi-FL=sym}
\eea
describe a local "Fermi-liquid" regime at the boundary. The impurity spin is totally screened
and its ground state is
non-degenerate: at $T=0$ the local boundary magnetization smoothly vanishes as $h_0 \to 0$.
\medskip

As follows from the Emery-Kivelson solution, the logarithmic divergence of the local susceptibility
(\ref{2H-log}) is 
weaker than the Curie law. This is a signature of non-Fermi-liquid character of the local response.
The logarithimic asymptotics of the boundary susceptibility (\ref{2H-log}) 
are the \emph{hallmark of Majorana physics at the edge of a single chain}.
As discussed below, these asymptotics
hold in the massive case as well when the quantun critical regime $|m| \ll T \ll \Gamma$
is considered.

\subsubsection{Disordered phase, $m>0$}\label{dis-phase}

As we already know, there are two boundary MZMs in the disordered phase. One of them
($\beta_2$) is completely decoupled whereas the other Majorana mode ($\beta_1$) hybridizes with
the bulk degrees of freedom. The behavior of the thermodynamic quantities at the edge 
in various temperature regimes will depend
on the ratio $m/\Gamma$, or equivalently, on the relation between two length scales --
the correlation length $\xi_m \sim v/m$ and the hybridization length $\xi_{\Gamma} \sim v/\Gamma$.
At $\xi_m \ll \xi_{\Gamma}$ ($m \gg \Gamma$) both boundary MZMs are well localized,
and one expects  
a Curie law for the local susceptibility,
$\chi_0 = C/T$, with a slightly reduced Curie constant ($C \lesssim 1$). In the other limit,
$\xi_m \gg \xi_{\Gamma}$ ($m \ll \Gamma$), hybridization effects are dominant. If
$m \ll T \ll \Gamma$, the mass gap can be ignored and the 
logarightmic %(2-channel Kondo)
regime (\ref{2H-log}) is seen. However, upon further decreasing the temperature,
$T \ll m \ll \Gamma $, a low-$T$ 
Curie behavior of the impurity reappears, albeit with a strongly reduced Curie constant.
\medskip

\begin{figure}[hbbp]
\centering
\includegraphics[width=2.5in]{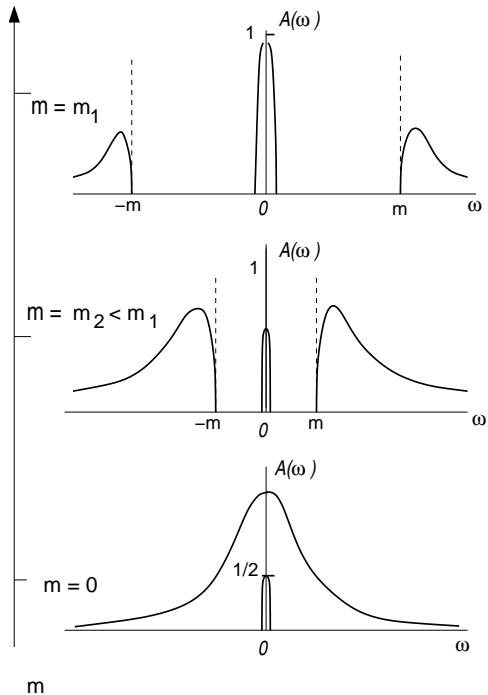}
\caption{\footnotesize Evolution of the spectral weight in the disordered phase ($m>0$) upon
decreasing $m$.
The lowest plot corresponds to $m=0$, where the continuum shoulders tend to $\omega\to0$. In this case the continuum spectral function  is to be understood as contributing only to the principal part when appearing in an integral, on top of which there is still a $\delta$-function contribution from $\omega=0$.
}\label{Aneg}
\end{figure}

With this qualitative picture in mind, let
us now turn to the spectral weight at $m>0$ given by Eq.~(\ref{A-m>0}).
At $h_0 = 0$,
its evolution on decreasing the mass $m$
is shown in Fig.~\ref{Aneg}. 
The low-frequency part of $A(\omega)$ 
contains a $\delta$-function singularity
\bea
A_{\rm sing} (\omega) = A_0 \delta(\omega),
~~~~A_0 = \frac{1}{2} \left( 1 + Z \right) =
\frac{1 + (\Gamma/4m)}{1 + (\Gamma/2m)} 
\label{A-singular}
\eea
where $Z$ is given by (\ref{Z}).
Its amplitude $A_0$ interpolates between the free-spin value $A_0 = 1$ in the strong localization
limit ($\Gamma/m \to 0$) 
and the "two-channel Kondo" value $A_0 = 1/2$ (see Eq.~(\ref{A-EK})) in the strong hybridization limit
($\Gamma/m \to \infty$). 
To calculate $m(h_0, T)$ at 
$h_0 \to 0$
we need to know the low-frequency part of  $A(\omega; h_0)$
at a small nonzero $h_0$. 
Using (\ref{cal-G<})
at $|\omega|, h_0 \ll m$ we obtain
\bea
A_{<}(\omega; h_0) %= - \frac{1}{\pi} \Im m~{\cal G}_<(\omega + \ri \delta) 
\simeq 
\frac{(\omega - 2h_0)^2} {4|\tilde{h}_0|} Z
\delta \left( \omega^2 - 4\tilde{h}^2 _0 \right), \nn
\eea
where 
$
\tilde{h}_0 = {h_0}\sqrt{Z}.
$
Here we only need the antisymmetric part of $A(\omega)$:
\bea
A^{\rm (as)}_{<} (\omega; h_0) = - \frac{1}{2} \sqrt{Z}
\left[  \delta (\omega - 2\tilde{h}_0 ) - \delta (\omega + 2\tilde{h}_0 )\right].
\label{A-antisym}
\eea
Substituting (\ref{A-antisym}) into (\ref{m-n-via-A}) we obtain 
\bea
m_{0;<} (h_0, T) = - \int_{-\infty}^{\infty} \rd \omega~
A^{\rm (as)}_{<} (\omega; h_0)
\tanh\frac{\omega}{2T}
= \sqrt{Z} \tanh \frac{\sqrt{Z}h_0}{T}.
\label{m-0Tneq0}
\eea
This expression yields the picture of a boundary spin
$\mu_{\rm eff} \s^z _0$, where $\mu_{\rm eff} = \sqrt{Z}$ is the effective magnetic
moment.
If the temperature is kept finite and $h_0 \to 0$, $m_{0;<}$ follows a Curie law,
$m_{0;<} = \chi_<(T) h_0$, where $\chi_<(T)$ is given by (\ref{CURIE}).
The zero-temperature value of $m_{0;<}$ %boundary magnetization 
displays the expected discontinuity at $h_0 \to
\pm 0$:
$m_{0;<} = \mu_{\rm eff} ~{\rm sgn}(h_0)$.
\medskip

\medskip

The high-frequency part of the spectral weight 
features a broad continuum of states at $\omega^2 > m^2$ with a non-singular behavior
at the thresholds $\omega = \pm m$: $A(\omega) \sim \sqrt{\omega \mp m}$.
To estimate the contribution of this frequency region 
to the boundary spin susceptibility
we proceed from the expression (\ref{cal-G>}) for 
${\cal G}_> (\omega)$,
find in the leading order in $h_0$ the antisymmetric part of the spectral weight
and, using formula (\ref{m-n-via-A}), obtain the high-frequency contribution to the
local susceptibility:
\be
\chi_> = \frac{4\Gamma}{\pi} \int_m ^{\infty} \rd \omega~ \tanh \left(\frac{\omega}{2T}\right) \cdot
\frac{\sqrt{\omega^2 - m^2}}
{\omega^2 (\omega^2 + \Omega^2 _1)},\label{chi>-general:m>0}
\ee
where
$
\Omega^2 _1 = \Gamma (2m + \Gamma) > 0.
$
At temperatures $T \ll m$ and arbitrary $\Gamma/m$,  ~$\chi_>(T)$ 
can be replaced by
its zero-temperature value 
\be
\chi_> (T=0) = \frac{4\Gamma}{\pi} \int_m ^{\infty}  \rd \omega~
\frac{\sqrt{\omega^2 - m^2}}
{\omega^2 (\omega^2 + \Omega^2 _1)}
\simeq \frac{4}{\pi (2m + \Gamma)} \ln \left( \frac{m+\Gamma}{m} \right).
\label{chi>-general:m>0-1}
\ee
%Here we replaced the square root in the integrand by $\omega$. This is accurate for $\Gamma/m\gg 1$, %while for $\Gamma/m\ll 1$ it only affects the numerical prefactor of the result which scales as 
%$\Gamma/m^2$. 
In the regime $m \gg \Gamma$ when the impurity spin is strongly localized,
the
high-frequency part of the impurity spectral weight gives only a small 
correction compared to the contribution of the already considered low-frequency part, $\chi_< (T)$,
the relative correction being 
of the order
of $T\Gamma/m^2$ for $T\ll m$, or of order $\Gamma/m$ for $T \gg m \gg \Gamma$.
\medskip

The situation changes in the strong hybridization limit, $m \ll \Gamma$. As the ratio $m/\Gamma$ tends to zero, the effective magnetic moment
$\mu_{\rm eff}$ 
%\to 0$ 
vanishes and, at the same time, 
the high-frequency contribution to the local susceptibility, Eq.~(\ref{chi>-general:m>0-1}),
logarithmically diverges. 
This means that at $m \ll \Gamma$
this contribution %of the incoherent, high-energy continuum of states
becomes dominant if $T > m$.
Indeed, replacing $\Omega_1$ by $\Gamma$ in (\ref{chi>-general:m>0})  we obtain:
\bea
\chi_>  (T)
&=&
~\frac{1}{T},  ~~~~~~~~~~~~~~~~~~T \gg \Gamma, \label{large-G-curie}\\
 &=&
\frac{4}{\pi \Gamma} \ln \frac{\Gamma}{\{ T,m \}}, ~~~~T \ll \Gamma,
\label{2H-chi}
\eea
where $\{ a,b\} \equiv {\rm max}(a,b)$.
We see that at 
$m \ll \Gamma \ll T$, the Curie-law behavior of the local susceptibility
is contributed by the incoherent, high-energy continuum of states.
At lower temperatures, 
$T,m \ll \Gamma$, ~$\chi_>$  follows the logarithmic two-channel-Kondo  
asymptotics~\cite{EK}, so that the total
local susceptibility 
is given by the sum
\be
\chi_0 (T) = \chi_< (T) + \chi_> (T) = \left( \frac{2m}{\Gamma} \right) \frac{1}{T}
+ \frac{4}{\pi \Gamma} \ln \frac{\Gamma} {\{T,m \}}.\label{ch-sum><}
\ee
Therefore,  at temperatures $T \ll m /\ln(\Gamma/m)$ a Curie regime with a small
Curie constant is recovered:
\be
\chi_0 (T) = \frac{C}{T}, ~~~C = \mu^2 _{\rm eff} \simeq  \frac{2m}{\Gamma}.
\label{lowest-T}
\ee
%$\chi_0 = C/T$, ~$C = 2m/\Gamma \ll 1$.
The high-temperature  and low-temperature Curie behaviors
of $\chi_0$ 
"sandwich" the intermediate-temperature ($m \ll T \ll \Gamma$) logarithmic asymptotics 
(\ref{2H-log}).
\medskip

Thus, at any \emph{finite} $m > 0$ (i.e. $Z\neq 0$)
there exists a well defined spin-1/2 degree of freedom localized at the open boundary
of the chain and characterized by the effective magnetic moment
$\mu_{\rm eff}$ which depends on the ratio $m/\Gamma$. The effective 
%magnetic 
moment
takes the value $\mu_{\rm eff} = 1$ at $\Gamma/m \to 0$ and decreases
upon increasing the ratio $\Gamma/m$, vanishing at criticality ($m=0$).
In the strong localization limit, $m \gg \Gamma$, the Curie law $\chi_0 \simeq 1/T$
is valid at any temperature.
Delocalization of the  impurity fermion $\beta_1$ across the whole chain in the critical state of the system is concomitant with the disappearance of the boundary spin-1/2 degree of freedom
($\mu_{\rm eff} \to 0$ as $m \to 0$). This is consistent with the emerging non-Curie,
logarithmic temperature dependence of the
local susceptibility, Eq.~(\ref{2H-chi}). 
Exactly at criticality ($m=0$), the local magnetization at $T=0$ is a non-analytic function of $h_0$\cite{EK}:
$
m(h_0) = (8h_0 /\pi \Gamma) \ln (\Gamma / h_0).
$
\medskip

%%%%%%%%%
\subsubsection{Effective moment of the Curie law: a probe of similarity of degenerate ground states}
It is useful to look at the effective magnetic moment $\mu_{\rm eff}$ from a different
perspective, using a Lehmann representation as in Eq.~(\ref{mu-lehmann}), where 
 $\mu_{\rm eff}$ was expressed as the 
matrix element of the operator $\sigma_0^z$ between the two degenerate ground states.
A renormalized but finite  local free-spin susceptibility (\ref{CURIE}) in the disordered phase reflects the fact that the two ground states only differ locally, that is, in a finite number of degrees of freedom. This shows that in this phase the origin of the exact spectral degeneracy is purely local, and accordingly, cannot be seen in measurements which only probe degrees of freedom far from the impurity. The vanishing of $\mu_{\rm eff}$ upon tuning to the critical point and in the ordered phase can be interpreted as an emerging %Anderson 
orthogonality between the ground states for $h_0=0\pm$. It is this orthogonality at the degeneracy point $h_0=0$, which protects the susceptibility from diverging in the ordered (topological) phase. 
\medskip

To put this result into a more general context, it is  useful to switch to the eigenbasis, in which the parity {operator} $P_S$ has a definite eigenvalue. Since $P_S$ commutes with $H$ irrespective of the value of the transverse field $h_0$, this basis is better adapted to discuss the crossing of the degeneracy point $h_0=0$. In the QIC model it is easy to prove that the ground states associated with transverse fields $h_0$ and $-h_0$ have opposite parity. This follows immediately from the fact that conjugation with $\sigma_0^x$ {($H \to \s^x _0 H \s^x _0, ~P_S \to \s^x _0 P_S \s^x _0$)} flips both the sign of $h_0$ and the parity operator. This implies that at $h_0=0$ the ground state is degenerate and that its parity flips, as $h_0$ is tuned across $0$.
\medskip

Let us now analyze a generic system in which the ground state is tuned to 
{a point where it becomes doubly degenerate or has exponentially small level splitting $\delta$},
while 
the {remaining} states are separated by a {finite} gap $\Delta$.  Let the two-dimensional ground state manifold be spanned by the orthogonal eigenstates $|\alpha\rangle$ and $|\beta \rangle$. At temperatures $\delta\ll T\ll \Delta$, the density matrix is simply proportional to  unity in this subspace, 
\bea
\rho= \frac{|\alpha\rangle\langle \alpha|+|\beta\rangle\langle \beta|}{2}. 
\eea
The  susceptibility of an observable $A$, {defined as the linear response} to its conjugate field, is easily calculated  to have the low temperature asymptotics of a Curie law,
\bea
\label{mueff}
\chi_A &=& \int_0^{1/T} d\tau \,\left[\langle A(\tau) A(0)\rangle - \langle A\rangle^2\right] 
= \frac{1}{T} \left[ |A_{\alpha\beta}|^2 + \frac{1}{4}(A_{\alpha\alpha}-A_{\beta\beta})^2\right],
\eea
where {$A_{\nu\nu'}= \la\nu| A|\nu' \ra$}. This {expression} is manifestly basis-independent when rewritten as {$\chi_A = \mu^2 _{\rm eff}/T$} with
\bea
\label{mueff_inv}
%\chi_A &=& \frac{\mu_{\rm eff}^2}{T},\quad {\rm with}\nn\\
\mu_{\rm eff}^2 =  \left[\frac{{\rm Tr}(A_{\gamma\eta})}{2}\right]^2 - {\rm Det}(A_{\gamma\eta})= \frac{(\lambda_1-\lambda_2)^2}{4},
 \eea
where $\lambda_{1,2}$ are the eigenvalues of the restriction of $A$ to the ground state manifold.
\medskip

In the quantum Ising model, {the} degeneracy occurs at $h_0=0$ and the local transverse susceptibility corresponds to the operator $A=\sigma^z_0$.  %Choosing 
{If $|\alpha\ra, |\beta\ra$ are chosen} to be the eigenstates $|+\rangle,|-\rangle$ of $\sigma_0^x$, {then} only 
the off-diagonal elements are nonzero, and one recovers Eq.~(\ref{mu-lehmann}) for $\mu_{\rm eff}$. Choosing instead the parity eigenbasis, $|P_S=\pm 1\rangle =\frac{1}{\sqrt{2}} (|+\rangle \pm |-\rangle)$,  only the diagonal elements can be nonzero by symmetry. 
{As follows from (\ref{mueff_inv}),}
in the disordered phase $\mu_{\rm eff}$ can assume any value $\leq 1$. 
However, anywhere in the ordered phase, $\mu_{\rm eff}$ becomes exponentially small in the system size, because of spontaneous symmetry breaking: The parity-even and -odd ground states are equal weight superpositions of the symmetry breaking ground states $|\pm \rangle$ (magnetization aligned or anti-aligned with $x$), with equal or opposite signs, respectively. This implies equality of the diagonal matrix elements in the parity basis, up to an exponentially small difference given by the matrix element $\langle - | \sigma_0^z| +\rangle$, which  connects the two symmetry breaking sectors.
\medskip

After the JW transformation, the symmetry-related degeneracy translates into the topological degeneracy of the Majorana Hamiltonian, with parity-even and -odd sectors being degenerate up to exponentially small perturbations. The latter feature is generic in topological phases, and essentially constitutes their defining property - topological protection: The action of any local observables, such as $A$,  restricted to the topologically degenerate manifold is that of a unit operator, up to corrections which are exponentially small in the system size. 
From this and formula (\ref{mueff_inv}) it follows immediately that the Curie weight $\mu_{\rm eff}^2$ is exponentially suppressed in a topological phase. This is the case even at exact degeneracy points, where the parity of the ground state switches (as happens, e.g., for $h_0=0$ in the fermionic version of the QIC in its ferromagnetic phase). We will see in Sec.~\ref{1DpwS} below a non-trivial example of a 1d p-wave superconductor, where the suppression of a Curie-like divergence of the low temperature charge susceptibility is found everywhere in the topological phase, independent of the fine-tuning of the local potential $\mu_0$ acting on an impurity site. As we will argue, the only generic exact ground state degeneracy occurs when the ground state switches parity within the topologically degenerate manifold. To bring about further degeneracies within the same parity sector  requires a high degree of fine-tuning. Such degeneracies will thus generically not be encountered upon moving along a single parameter family of Hamiltonians, such as varying a local potential or field.
%%%%%%%%%%%%%%%%%%%%%%
\bigskip

\subsubsection{Ordered phase, $m<0$}

\begin{figure}[hbbp]
\centering
\includegraphics[width=2.5in]{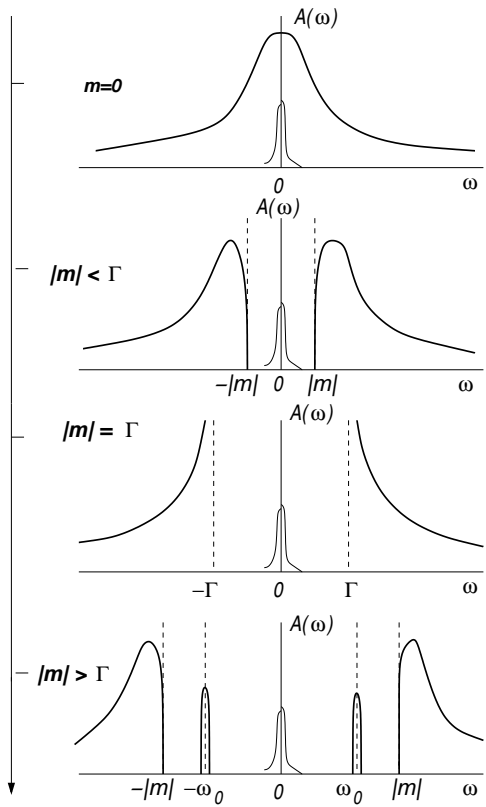}
\caption{\footnotesize Evolution of the spectral weight in the ordered phase ($m<0$) upon
increasing $|m|$. As the bulk gap $|m|$ crosses $\Gamma$, a pair of discrete subgap states emerges.}\label{Apos}
\end{figure}

The physical picture emerging in the ordered phase ($m<0$) is qualitatively different from the disordered phase.
As follows from (\ref{A-relation}), the completely decoupled MZM  $\beta_2$
gives a contribution $(1/2) \delta(\omega)$ to the spectral weight
which remains intact
upon varying the ratio
$|m|/\Gamma$ and %, of course, and 
is immune against application of external fields.
{The transformation of the incoherent high-frequency background on
increasing the mass gap $|m|$ is shown in 
Fig.~\ref {Apos}}.
As long as $|m| < \Gamma$, the behavior of $A(\omega)$ is
qualitatively the same as in the disordered phase. At $|m|=\Gamma$
the spectral weight acquires new features. Precisely at this point
the thresholds at $\omega = \pm |m|$
transform from non-singular to singular: $A(\omega) \sim 1/\sqrt{\omega \mp |m|}$.
This feature comes together with the "birth" of a particle
%This singularity is a precursor of a "particle creation" 
in the spectrum:
at $|m| = \Gamma + 0$
two symmetric $\delta$-function peaks, $\delta(\omega \mp \omega_0)$,
emerge  just below the thresholds. 
They belong to new levels with energies  inside the gap, which 
split from the incoherent continuum of states. %emerge 
They are mainly due to the hybridization between
two zero modes -- the impurity $\beta_1$-Majorana fermion and
the zero mode $\gamma_0$, which is present in the spectrum of the Dirac Hamiltonian at 
$m<0$. These discrete sub-gap states exist provided that the localization of $\gamma_0$ is strong enough (the condition $|m| > \Gamma$).
The effective Hamiltonian describing this splitting in the limit $|m|\gg \Gamma$
simply coincides with the $\lambda_0$-mixing term in the model (\ref{maj-ham-final12}):
$H' = -\ri \lambda_0 \beta_1 \gamma_0$, where $\lambda_0 \sim \sqrt{\Gamma |m|} \sim \omega_0$.
Notice that as soon as $|m|>\Gamma$
the spectra at the thresholds $\omega = \pm |m|$ return to a
non-singular form. 
\medskip

{The main features of the spectral weight $A(\omega)$ in the strong localization limit, 
$|m| > \Gamma$, shown schematically in the last plot of Fig.~\ref {Apos}, are in a qualitative agreement
with the results of a recent numerical work \cite{korytar}, where the spectrum of a resonant level
attached to the edge of a 1D triplet superconductor was calculated. However, local thermodynamic properties
of this system have not been addressed in Ref.~\onlinecite{korytar}.}

\medskip

\medskip

Let us determine the effect of these subgap states on the susceptibility of the impurity spin.
At $h_0 \ll \omega_0$ the positions
of the peaks $\delta (\omega \mp \omega_0)$ change only by an amount $\sim O(h^2_0)$, which 
can be safely neglected.
As a result  %, at small $h_0$ 
the antisymmetric part of the low-frequency ($\omega^2 < m^2$) spectral weight $A_<(\omega)$ is simply proportional to
$h_0$:
\bea
A^{\rm (as)}_< (\omega) = - \frac{2h_0}{\omega_0}
\theta(|m| - \Gamma) \left(\frac{|m|-\Gamma}{2|m|-\Gamma} \right)
\left[ \delta(\omega - \omega_0)
- \delta (\omega + \omega_0)  \right], ~~~m < 0. \label{A-antisym1}
\eea
Accordingly,  $\chi_<$
is equal to
\be
\chi_<(T) = \theta(|m| - \Gamma) \frac{4}{\omega_0}\left(\frac{|m|-\Gamma}{2|m|-\Gamma} \right)
\tanh \frac{\omega_0}{2T}. \label{chi<0}
\ee
At $T \gg \omega_0$ the $\delta$-function contribution
to the susceptibility follows a Curie law
$
\chi_<(T) = 
\bar{\mu}^2 _{\rm eff}/T, % \label{chi<0+}
$
where the effective boundary magnetic moment 
$
\bar{\mu}_{\rm eff} = \left[ 2(|m|-\Gamma)/(2|m|-\Gamma)\right]^{1/2}
$
varies from $\bar{\mu}_{\rm eff} = 1$ at $|m|/\Gamma \gg 1$ to $\bar{\mu}_{\rm eff} = 0$ as
$|m| \to \Gamma + 0$. 
As follows from (\ref{chi<0}), at lower temperatures,
$T \ll \omega_0$, %the boundary spin gets fully screened, and
$\chi_<(T)$
crosses over to a constant, temperature independent
value
\bea
\chi_< (T = 0) = 4\theta(|m| - \Gamma) \left(\frac{|m|-\Gamma}{2|m|-\Gamma} \right)
\frac{1}{\sqrt{\Gamma (2|m| - \Gamma)}}.
\label{chi<0-}
\eea
In particular, at 
strong inequality, $|m| \gg \Gamma$, we have
\be
\chi_< (T=0) = \frac{2}{\sqrt{2\Gamma |m|}}, \label{chi:m<0}
\ee
for which we will give a simple heuristic explanation below.
\medskip

Estimating 
the effect of the high-frequency spectral function
on the susceptibility $\chi_0$, we find  that the 
expression for $\chi_>$ for the ordered phase 
is obtained similarly as in the disordered phase:
\bea
\chi_> (T) 
= \frac{4\Gamma}{\pi} \int_{|m|}^{\infty} \rd \omega~\tanh \left( \frac{\omega}{2T} \right)
\frac{\sqrt{\omega^2 - m^2}}{\omega^2 (\omega^2 - 2|m|\Gamma + \Gamma^2)}.
\label{chi>-thresh}
\eea
Using (\ref{chi>-thresh})
one can easily check that at the threshold $|m|=\Gamma$ where the spectral weight
becomes singular, the local susceptibility remains analytic. 
Turning to the strong localization regime, $|m| \gg \Gamma$, we find that
the $T=0$ value of $\chi_>$ is given by
\bea
\chi_>(T=0) = \frac{4\Gamma}{\pi}\int_{|m|}^{\infty} \rd \omega~\frac{\sqrt{\omega^2 - m^2}}{\omega^2
(\omega^2 - \omega^2 _0)}
\simeq \frac{4\Gamma}{\pi}\int_{|m|}^{\infty} \rd \omega~\frac{\sqrt{\omega^2 - m^2}}{\omega^4}
\sim \frac{\Gamma}{m^2}, \nn %\label{chi>:large-m}
\eea
where we have used the fact that $\omega_0 = \sqrt{2\Gamma |m|}\ll |m|$.
In the same limit $\chi_<$ is given by (\ref{chi:m<0}). Therefore
$
{\chi_>}/{\chi_<} \sim ( {\Gamma}/{|m|} )^{3/2} \ll 1.
$
\medskip

Thus we conclude that in the strong localization limit ($|m| \gg \Gamma$),
%the state of the boundary impurity in the ordered phase is nondegenerate.
the dominant contribution to the local susceptibility $\chi_0 = \chi_< + \chi_>$ comes from
its low-frequency part $\chi_<$, Eqs.~(\ref{chi<0})--(\ref{chi:m<0}). 
\medskip

The result (\ref{chi:m<0}) can be understood as follows. The impurity spin $\s^x _0$ couples by an exchange
interaction $\tilde{J}$ to the spin $\s^x _1$ located at the boundary site $j=1$ 
of the semi-infinite QIC:
$H_{01} = - \tilde{J} \s^x _0 \s^x _1$. In the ordered phase $\s^x _1$ acquires an expectation value,
and in the leading order the impurity spin experiences a local magnetic field
$h_x = \tilde{J} \la  \s^x _1\ra$. Adding a local transverse field ${\bf h}_0 = (0,0, h_0)$, we write the
total energy of the impurity spin as
$
H_{01} = - h_x \s^x _0 - h_0 \s^z _0.
$
In the limit $h_0 \to 0$, the transverse spin susceptibility
is equal to
$
\chi_0 = 1/h_x \sim 1/ \sqrt{\Gamma}\la \s^x _1 \ra,
$
where $\Gamma \sim \tilde{J}^2 /v$. The crucial point is that the scaling 
(i.e., the mass dependence)
of the boundary magnetization
in a slightly non-critical (ordered) QIC is different from that in the bulk\cite{ghoshal}:
$
\la \s^x \ra_{\rm bound} \sim \sqrt{|m|}.
$
This leads to
$
\chi_0 \sim 1 / \sqrt{\Gamma |m|}.
$
\medskip

It remains to  consider the regime $|m| \ll \Gamma$, which is entirely contributed
by the high-frequency part of the impurity spectral weight. In this regime the role of the mass gap 
$|m|$ reduces to an infrared cutoff in the logarithmic temperature dependence of the local susceptibility,
and the sign of the mass is completely unimportant. Therefore, replacing $m$ by $|m|$, the formulas (\ref{large-G-curie}),
(\ref{2H-chi}) of the preceding subsection can be used in this case as well:
\bea
|m| \ll \Gamma \ll T: && \chi_0 = \frac{1}{T}, \label{m<0:high-T-EK}\\
|m|, T \ll \Gamma: && \chi_0 = \frac{4}{\pi \Gamma} \ln \frac{\Gamma}{\{ T, |m|\}}.
\label{m<0;low-T-EK}
\eea
The zero-temperature local susceptibility is given by
\be
\chi_0(T=0; |m|\ll \Gamma) = \frac{4}{\pi \Gamma} \ln \frac{\Gamma}{|m|}.
\label{m<0:T=0-chi}
\ee

The results obtained in this subsection lead us to the conclusion that the state of the boundary impurity in the ordered
phase is nondegenerate.

\medskip

The asymptotic behavior of the impurity spin susceptibility in different temperature regimes
is summarized in the tables below:

\medskip

{\sl Disordered phase: $m>0$}
\bea
m \gg \Gamma: && \chi_0 = \frac{1}{T} ~~~~~~~~~~~{\rm at~all~~}T,\nn\\
m \ll \Gamma: && \chi_0 = \frac{1}{T} ~~~~~~~~~~~{\rm at} ~~T \gg \Gamma, \nn\\
&& \chi_0 = \frac{4}{\pi \Gamma} \ln \frac{\Gamma}{T} ~~~
{\rm at} ~~ \frac{m}{\ln (\Gamma/m)}\ll T \ll \Gamma,
\nn\\
&& \chi_0 = \left( \frac{2m}{\Gamma} \right)\frac{1}{T} ~~{\rm at} ~~T \ll \frac{m}{\ln (\Gamma/m)}.
\label{summary1}
\eea
\bigskip

{\sl Ordered phase: $m<0$}
\bea
m \gg \Gamma: && \chi_0 = \frac{1}{T} ~~~~~~~~~~~~{\rm at} ~~T \gg \sqrt{\Gamma |m|},\nn\\
&& \chi_0 = \frac{2}{\sqrt{2\Gamma |m|}} ~~~~{\rm at} ~~T \ll \sqrt{\Gamma |m|},\nn\\
m \ll \Gamma: && \chi_0 = \frac{1}{T} ~~~~~~~~~~~~{\rm at} ~~T \gg \Gamma, \nn\\
&& \chi_0 = \frac{4}{\pi \Gamma} \ln \frac{\Gamma}{T} ~~~~{\rm at} ~~
|m| \ll T \ll \Gamma, \nn\\
&& \chi_0 = \frac{4}{\pi \Gamma} \ln \frac{\Gamma}{|m|} ~~~{\rm at} ~~
T \ll |m|. \label{summary2}
\eea

\medskip

{
To conclude our discussion of a boundary impurity in the topological phase of a wire,
let us consider a more general situation in which}
%More generally, in the topological phase of the wire, we can describe 
the end of the wire is essentially described by an odd number $(2N-1)$ Majorana modes, $c_{2,...,2N}$, 
{with} the remaining degrees of freedom {being treated} as sufficiently strongly gapped and therefore irrelevant for the low energy sector.  The impurity is again described by two Majoranas, $c_{0}$
and $c_1$. The most general quadratic Hamiltonian then takes the form
\bea
{\cal H} = \ri \sum_{jk} A_{jk} c_j c_k, 
\eea
where $A_{jk}$ is a real, antisymmetric $(2N+1)\times (2N+1)$ matrix.
\medskip

Since the number of Majoranas is odd and the spectrum has a particle-hole symmetry, the presence of at least one exact zero mode is guaranteed. We now show that all other levels are generically repelled from zero. This is most conveniently seen from the characteristic polynomial of the matrix $\ri A$, 
\bea
p_{\ri A}(E) \equiv {\rm det}(\ri A-E) = \sum_{i=1}^{N} c_i(A) E^{1+2i},
\eea
where only odd powers of the energy variable $E$ appear with coefficients $c_i(A)$.
In order to find a fermionic level at $E=0$ further to the guaranteed Majorana zero mode, the characteristic polynomial must have a triple zero at $E=0$. This requires $c_1(A)=0$, which can be expressed in terms of the diagonal minors of $A$,
\bea
c_1(A) = (-1)^{N+1} \sum_{i=0}^{2N} {\rm det} (\hat A_{ii}) = (-1)^{N+1} \sum_{i=0}^{2N} [{\rm Pf}(\hat A_{ii})]^2 =0,
\eea
where $\hat A_{ii}$ is the matrix with the $i$'th row and the $i$'th column eliminated, and ${\rm Pf}(\hat A_{ii})$ is its Pfaffian.
Due to the antisymmetry of $A$, the minors ${\rm det}(\hat A_{ii})$ are squares of Pfaffians, which must all vanish simultaneously for any other level to cross $E=0$. It is clear that it requires a high degree of fine-tuning to circumvent the level repulsion from the guaranteed zero mode. The only obvious way to achieve this is by completely decoupling  one or several sites from the rest and driving a zero-crossing in that decoupled part of the system. However,  a generic variation of local parameters at the impurity or at the end of the chain will  not lead to an extra zero crossing.

%%%%%%%%%%%%%%%%%%%%%%%%%%%

\subsubsection{Qualitative difference of the boundary response in topologically ordered and disordered phases}

As discussed above, the local %boundary 
response of the system to a boundary transverse magnetic
field is qualitatively different in the topologically disordered and ordered phases:
in the disordered phase the response function $\chi_0 (T)$ follows a singular
Curie asymptotics in the $T \to 0$ limit, whereas in the ordered phase it is finite.
It is instructive to look again at this
qualitative difference by approaching it from the "polarization-loop" representation of $\chi_0 (T)$, 
Eq.~(\ref{chi-loop-gen}).
\medskip

Using the explicit form of 
the Majorana Green's functions $D_{aa}(\vare_n)~(a=1,2)$ at $h_0 = 0$ (see Appendix \ref{derivGF})
one represents $\chi_0 (T)$ as a sum over Matsubara frequencies
\be
\chi_0 (T) = 4T \sum_{\vare_n}\frac{1}{\Delta(\vare_n)}, ~~~~
\Delta(\vare_n) = \vare^2 _n + \Gamma \left( \sqrt{\vare^2 _n + m^2} - m\right). \label{chi-fin}
\ee
At $T \to 0$ the sum in (\ref{chi-fin}) transforms to the integral
\be
\chi_0 (0) = \frac{4}{\pi} \int_0 ^{\infty} \frac{\rd \vare}{ \Delta(\vare)}.
\label{chi:T=0}
\ee
For an isolated spin ($\Gamma = 0$) $\Delta(\vare) = \vare^2$ and the integral in
(\ref{chi:T=0}) diverges at the lower limit, implying that 
the Matsubara
sum in (\ref{chi-fin}) cannot be replaced by the integral (\ref{chi:T=0}). Doing the
Matsubara sum leads to
the standard Curie law:
\bea
\chi_0 (T) = 4T \sum_{\vare_n}\frac{1}{\vare^2 _n} =
\frac{8}{\pi^2 T} \sum_{n\geq 0} \frac{1}{(2n+1)^2} 
= \frac{1}{T}. \label{curie}
\eea

In spite of its simplicity, this result is quite noteworthy. It indicates that,
for a boundary impurity spin in a QIC, the response of the system to the local transverse
magnetic field $h_0$ will follow a singular Curie behavior
\be
\chi_0 (T\to 0) = \frac{C}{T}, ~~~0 < C < 1, \label{Curie-general}
\ee
only if $\Delta (\vare) \sim \vare^2$ as $\vare \to 0$. If, on the other hand,
$\Delta (\vare) \to {\rm const} \neq 0$ as $\vare \to 0$, the local boundary response
will be non-singular, and the limit $\lim_{T\to 0} \chi_0 (T)$ will be finite. In such a situation,
the ground state of the impurity spin is non-degenerate.
The arguments we will give below
unambiguously indicate that, irrespective of the magnitude of
the ratio $\Gamma/|m|$, the non-singular behavior of $\chi_0 (T)$ in the limit $T \to 0$
is entirely due to the presence of the boundary zero Majorana mode in the bulk spectrum of the
QIC at $m<0$.
\medskip

The contribution of the boundary MZM
can be singled out by representing $\Delta(\vare_n)$
as follows:
\be
\Delta(\vare_n) = \vare^2 _n  +
\Gamma ( \sqrt{\vare^2 _n + m^2} - |m| )
+ 2 \Gamma |m| \theta(-m).\label{Delta1}
\ee
Consider now
the region of small frequences,  $|\vare_n| \ll |m|$ (which automatically implies that
$T \ll |m|$) in which case
\be
\Delta(\vare_n) \simeq \left( 1 + \frac{\Gamma}{2|m|} \right)\vare^2 _n 
+ 2 \Gamma |m| \theta(-m). \label{Delta2}
\ee
It then %immediately 
follows from (\ref{Delta2}) that in the disordered phase ($m>0$), when
the second term in the r.h.s. of (\ref{Delta2}) is absent,
$\Delta(\vare_n) \sim \vare^2 _n $  and
the susceptibility maintains its singular Curie form (\ref{Curie-general}) with the renormalized
Curie constant:
\be
C = \mu^2 _{\rm eff} = \frac{2m}{\Gamma + 2m}. \label{C}
\ee 
Obviously, the results of Eqs.~(\ref{Curie-general},\ref{C}) are contributed by the low-frequency
$(\omega^2< m^2)$ part of the impurity spectral weight $A(\omega)$, while the contribution of the
high-frequency continuum ($\omega^2 > m^2$) represents a relatively small correction.
\medskip

On the other hand, in the ordered (topological) phase ($m<0$) 
$\Delta(\vare_n) \to 2 \Gamma |m| ~\neq ~0$ as $\vare_n \to 0$.
So, the presence of the boundary MZM in the spectrum at $m<0$ is the ultimate reason 
\emph{why}
the local response of the topologically ordered phase
is non-singular. This reflects the fact that, in the thermodynamic limit, coupling locally to a boundary degree of feedom does not allow to switch the topological sector, or the sector of spontaneously broken symmetry. Thus the linear response remains non-singular even at points where there is a spectral degeneracy between the different sectors. 
\medskip

However, this argument does not always imply that the finite value of $\chi_0$ in the
topologically ordered phase will be contributed by the region of small
$\vare$ where the expansion (\ref{Delta2}) is valid. Such an approximation is valid
in the limit
of strong localization, $|m| \gg \Gamma$. The sum over $\vare_n$ will then mostly be  contributed by frequencies $|\vare| \sim \sqrt{\Gamma |m|} \ll |m|$. In this case
we can approximate $\Delta(\vare)$ by the expression
\be
\Delta (\vare) \simeq \left( 1 + \frac{\Gamma}{2|m|} \right)\vare^2 + 2\Gamma |m|
\simeq \vare^2 + 2\Gamma |m|,
\label{expan2}
\ee
and estimate $\chi_0(0)$ 
\bea
\chi_0 (T=0) &=& \frac{4}{\pi} \int_0 ^{\vare_{\rm max}} 
\frac{\rd \vare}{\vare^2 _n 
+ 2 \Gamma |m|} = \frac{2}{\sqrt{2\Gamma |m|}}, ~~~m<0.
\eea
As follows from the structure of the impurity spectral weight $A(\omega)$, in the strong
localization limit the quantity $\omega_0 = \sqrt{2\Gamma |m|}$ represents a characteristic
energy scale of a subgap bound %(\AN{perhaps better: pairing}) 
state
of the impurity
Majorana $\beta_1$ and the boundary zero mode $\gamma_0$. The contribution of the high-frequency
continuum to $\chi_0 (0)$ is subdominant.
\medskip

In the strong hybridization limit, $\Gamma \gg |m|$, the situation is different. Let us first
estimate the contribution of small frequencies, $\vare^2 \ll m^2$. From (\ref{expan2})
we have
\bea
\Delta (\vare) 
\simeq \frac{\Gamma}{2|m|} (\vare^2 + 4 m^2). \label{expan3}
\eea
The contribution of the region $\vare^2 < m^2$ to $\chi_0 (0)$ in (\ref{chi:T=0})
is easily estimated:
\bea
\chi_< (0) = \frac{8|m|}{\pi\Gamma} \int_0 ^{\sim |m|} \frac{\rd \vare}{\vare^2 + 4m^2}
\simeq \frac{4}{\pi \Gamma}.
\label{chi<01}
\eea
At this point we should recall that at temperatures $T < |m| \ll \Gamma$
the high-frequency continuum of local states provides a %contributes to $\chi_0 (0)$
logarithmically enhanced contribution to the susceptibility
\be
\chi_> (0) \simeq \frac{4}{\pi \Gamma} \ln \frac{\Gamma}{|m|}, \label{2-kondo-m}
\ee
the mass gap $|m|$ serving 
as the infrared cutoff of the logarithm. 
Since within the logarithmic accuracy, a large logarithm $\ln (\Gamma/|m|)$
is admitted to 
have a relatively small correction $\sim O(1)$, 
in the strong hybridization limit $\chi_> (0)$ logarithmically dominates $\chi_<(0)$.
However, the most important fact here is that, contrary to the situation in the disordered phase,
$\chi_< (T=0)$ is \emph{finite} at $m<0$. This is due to the presence of the boundary MZM, as
is clearly seen from formulae (\ref{expan3},\ref{chi<01}). 
\medskip

Thus, the role of the boundary MZM in the topological phase is clear. Hybridization
with the bulk MZM %opens a gap in % it 
suppresses
the low-frequency part of the fluctuation spectrum of the impurity spin and renders the
local response non-singular, as illustrated in Table.~\ref{table:MainIllustration}. Therefore, the non-singular zero-temperature limit of the local response function $\chi_0$ of an impurity at the edge of a chain
serves as an indication of the existence of a boundary
Majorana zero mode in the topologically ordered phase.

%%%%%%%%%%%%%%%%%%%%%%%%%%%

\section{Impurity in the bulk of a quantum Ising chain}\label{bulk}

In this section we consider a zero-field impurity in the \emph{bulk (rather than the edge)} of a non-critical QIC. As
we have shown in Sec.~\ref{reduction}, this model reduces to a problem of two semi-infinite
QICs coupled to the impurity spin at the boundary. We will assume that the chains $a=1,2$
are identical ($v_1 = v_2$, $m_1 = m_2$) but characterized by independent nonzero
hybridization constants $\Gamma_1$ and $\Gamma_2$. At $\Gamma_ 1 = \Gamma_2$ %we have 
the model is equivalent %of a
to a spinless, semi-infinite Peierls insulator (PI) chain with a boundary fermionic $d$-level,
as we show in Appendix \ref{peierls}.
In the PI model, $\chi_0$  also 
describes the local compressibility at the impurity site. The role of the mass $m$
is played by the difference  between the alternating hopping amplitudes of the Peierls chain,
{$t_{\pm} = t \pm \Delta$}.
%It is positive, and the PI is non-topological,  if the first  two sites  next to the impurity form a %more strongly coupled dimer.
{For a positive $\Delta$ the first  two sites  next to the impurity form a more strongly coupled dimer, and 
the corresponding massive phase of the PI is non-topological. In the opposite case, $\Delta < 0$,
the ground state of the PI is topologically degenerate and supports boundary zero modes.}
\medskip

We might as well consider an impurity
located at a domain wall separating the ordered ($x>0$) and disordered ($x<0$) phases of a QIC. 
For topological reasons, to understand the low-energy sector, we can formally take the limit %of %infinitely large mass 
$m \to + \infty$ at $x \to -\infty$. Thus, in the low-energy limit,
%Then %it turns out that 
this case reduces to
the already considered problem of a single semi-infinite QIC in the ordered phase
($m<0$) with a boundary impurity. 
\vskip 0.2 truecm

For a bulk classical spin the local susceptibility
can be calculated for arbitrary $\Gamma_1$ and $\Gamma_2$ using the general formula (\ref{G-G1}) for $G(\vare_n, h_0)$ 
and the rules of analytic continuation (\ref{anal-cont}). 
Since at $h_0 = 0$ the impurity Majorana fermions $\beta_1$ and $\beta_2$ are decoupled,
the resulting model represents a direct sum of two semi-infinite KM chains
(\ref{not1-H}) (see Fig.~\ref{folded}). Accordingly, 
the spectral weight 
of the impurity center in such a model is given by 
\be
A(\omega) = A_1 (\omega) + A_2 (\omega) - \delta(\omega) =
\frac{1}{2}\left[A_{f1}(\omega) + A_{f2} (\omega)\right], ~~~(h_0 = 0),
\label{bulk-A-sum}
\ee
where 
$
A_a (\omega) = (1/2)\left[ \delta(\omega) + \frac{1}{2} A_{fa} (\omega)\right]
$
are the spectral weights of isolated semi-infinite QICs labeled by $a=1,2$. 
However, for $h_0 \neq 0$ the additive structure of $A(\omega)$, given by (\ref{bulk-A-sum}),
is no more valid.
\medskip

Using the definition (\ref{bulk-A-sum}) and formulae 
(\ref{A-m>0}), 
(\ref{A-m<0}), at $h_0 = 0$ we obtain

\bea
m>0: && A(\omega) ~=~ \frac{1}{2} (Z_1 + Z_2) \delta(\omega) + \theta(\omega^2 - m^2)
\frac{1}{2\pi} \sum_{a=1,2} \frac{\Gamma_a \sqrt{\omega^2 - m^2}}{|\omega|
(\omega^2 + 2\Gamma_a m + \Gamma^2 _a)};\label{bulk:m>0}\\
m<0: && A(\omega) ~=~ \frac{1}{2} \sum_{a=1,2}
\theta(|m|- \Gamma_a) \left( \frac{|m|- \Gamma_a}{2|m| - \Gamma_a}  \right)
\left[ \delta(\omega - \omega_a) + \delta (\omega + \omega_a) \right]\nn\\
&& ~~~~~~~~+ ~\theta(\omega^2 - m^2)\frac{1}{2\pi} \sum_{a=1,2} \frac{\Gamma_a \sqrt{\omega^2 - m^2}}{|\omega|
(\omega^2 - 2\Gamma_a |m| + \Gamma^2 _a)},\label{bulk:m<0}
\eea
where
\bea
Z_a &=& \frac{2m}{2m + \Gamma_a}, ~~~~~~~~~~~~~~~~~~~~~(m>0),\nn\\
\omega_a &=& \sqrt{\Gamma_a (2|m| - \Gamma_a)} < |m|~~~~(m<0).\nn
\eea

\subsection{Topologically disordered phase, $m>0$}

In the disordered phase ($m>0$), the low-frequency part
of the local spectral weight $A(\omega; h_0 =0)$ is contributed by the $\delta(\omega)$
singularity in the r.h.s. of (\ref{bulk:m>0}).
The local susceptibility follows the Curie law 
$\chi_{<} = \mu^2 _{\rm eff}/T$.
The effective magnetic moment of the
impurity  
\be
\mu_{\rm eff} = \sqrt{Z_1 Z_2} \label{mu-eff-bulk}
\ee
interpolates between the values $\mu_{\rm eff}
= \sqrt{Z_1}$ at $\Gamma_2 = 0$ (single semi-infinite QIC) and $\mu_{\rm eff} = Z_1$ at $\Gamma_2 = \Gamma_1$ (equivalent to the PI chain).
At $\Gamma_1, \Gamma_2 \neq 0$,  on approaching the criticality ($m \to 0$)
%in the strong hybridization limit for both chains, $m \ll \Gamma_1, \Gamma_2$,
%for two identical disordered chains ($\Gamma_1 = \Gamma_2$), in the limit
%$m \to 0$, 
the Curie constant scales as 
\be
C = \mu^2 _{\rm eff} \sim \frac{m^2}{\Gamma_1 \Gamma_2}.
\label{m^2-scaling} 
\ee
It thus approaches zero much faster than in the case of a single chain with a boundary
impurity ($\Gamma_2 = 0$), where $C$ scales as $C\sim m$.
\medskip

The contribution of the high-frequency part of the spectral weight to the local susceptibiliy
is given by
\bea
\chi_{>} = \frac{4}{\pi} \int_m ^{\infty} \rd \omega~\frac{\sqrt{\omega^2 - m^2}}{\omega^2}
\tanh \left( \frac{\omega}{2T} \right)
\frac{\Gamma_1  (\omega^2 + \Gamma_2 m)
 + \Gamma_2  (\omega^2 + \Gamma_1 m)}{(\omega^2  +\Omega^2 _1)(\omega^2 + \Omega^2 _2)},
\label{A-as-m>0}
\eea
where $\Omega_a ^2 = \Gamma_a (2m + \Gamma_a)$.
As for a single QIC, in the strong localization limit, $|m| \gg \Gamma_{1,2}$,
$\chi_{>}$ in (\ref{A-as-m>0}) leads to small
corrections to the low-frequency contribution to $\chi_{<}$.
\medskip

The frequency region $\omega^2 >m^2$ becomes important when 
the impurity state is strongly hybridized with at least one of the two chains.
To analyze such cases, it is convenient to transform $\chi_{>}$ in (\ref{A-as-m>0}) to an equivalent form
\bea
\chi_{>} 
= \frac{\Gamma_1 \chi_{>}^{(1)} - \Gamma_2 \chi_{>}^{(2)}}{\Gamma_1 - \Gamma_2},
\label{chi>12}
\eea
where 
$\chi_{>}^{{a}}~(a=1,2)$ 
are the high-frequency contributions to the susceptibilities
of isolated QICs, Eqs. (\ref{chi>-general:m>0}), already discussed in Sec.~\ref{semi} B. 
\medskip

Suppose that $\Gamma_1 \gg \Gamma_2$. We consider first the case $\Gamma_2 \ll m \ll \Gamma_1$.
It describes a situation when the impurity Majorana fermion $\beta_1$ is strongly hybridized
with the bulk excitations of the first chain, whereas the fermion $\beta_2$ is strongly
localized. So, the impurity MZM $\beta_2$ can be approximately regarded as completely decoupled from
both chains. %Indeed, as
It follows from (\ref{chi>-general:m>0-1}), at $\Gamma_2 \ll m \ll \Gamma_1$
$
~\chi_{>}^{(2)} \simeq \Gamma_2 /m^2.
$
Then, as one expects, up to small relative corrections $\sim \Gamma_2/\Gamma_1$ and $(\Gamma_2/m)^2$, the total susceptibiity $\chi_0$ coincides with  $\chi^{(1)}_0$, i.e. the local susceptibility of an isolated
semi-infinite first chain ($a=1$) in the strong hybridization regime.

\medskip

Consider now the case when both chains are in the strong hybridization regime, %takes place for both $%chains, 
$m \ll \Gamma_1, \Gamma_2$.
In the temperature range $m \lesssim T \ll \Gamma_1, \Gamma_2$, using the logarithmic asymptotics 
(\ref{2H-chi}) for $\chi_{>}^{(1,2)}$, we arrive at a temperature independent susceptibility
(a "local-Fermi-liquid" regime):
\bea
\chi_0 \simeq \chi_{>}= %\lim_{\Gamma_1 \to \Gamma_2}
\frac{4}{\pi(\Gamma_1 - \Gamma_2)} \ln \frac{\Gamma_1}{\Gamma_2}. %= \frac{4}{\pi \Gamma}
\label{1=2:const-chi}
\eea
For two chains  coupled symmetrically to the impurity, $\Gamma_1 = \Gamma_2 \equiv \Gamma$,
the result (\ref{1=2:const-chi}) becomes
$
\chi_0 = 4/\pi \Gamma.
$
According to (\ref{m^2-scaling}), at $m \ll \Gamma$ the effective Curie asymptotics 
(contained in $\chi_<$)
is 
$(m/\Gamma)^2 /T$.
Comparing this to (\ref{1=2:const-chi}) we find that
\bea
\chi_0 &\simeq&  \frac{4}{\pi \Gamma}, ~~~~~~~~~~~~{\rm if}~~\frac{m^2}{\Gamma} \ll T \ll m, 
\label{cont}\\
\chi_0 &\simeq& \left(\frac{2m}{\Gamma}\right)^2 \frac{1}{T}, ~~~
{\rm if}~~ T \ll \frac{m^2}{\Gamma}. \label{2curie}
\eea
Thus, also in the channel-symmetric case (or equivalently, in a PI chain),
there exists a re-entrant crossover between
two Curie regimes, $\chi_0 = 1/T$ at $T \gg \Gamma$ and $\chi_0 = (2m/\Gamma)^2 /T$
at $T \ll m^2/\Gamma$, separated by a temperature independent 
susceptibility plateau (\ref{cont}).

%%%%%%%%%%%%%%%%%%%%%%%
\subsection{Topologically ordered phase, $m<0$}

\begin{figure}[hbbp]
\centering
\includegraphics[width=3.9in]{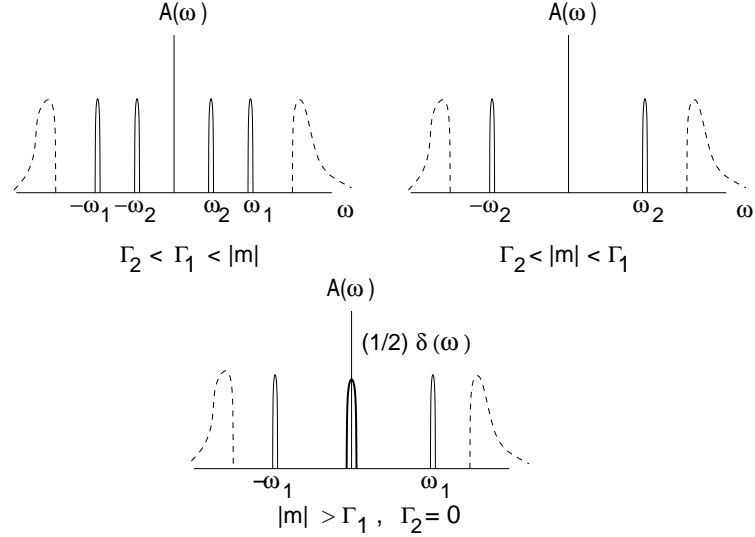}
\caption{\footnotesize The spectral weight of an impurity spin coupled to
two quantum Ising chains in the absence of a local field ($h_0 = 0$).}
\label{gamma12}
\end{figure}

In the ordered phase ($m<0$), the spectral weight $A(\omega; h_0 =0)$ at $\omega^2 < m^2 $
is given by the first term in the r.h.s. of (\ref{bulk:m<0}).
The subgap peaks shown in Fig.~\ref{gamma12} describe bound states between the impurity
Majorana fermions $\beta_1, \beta_2$ and the MZMs of the bulk spectra of the
corresponding chains. Depending on the relation between the parameters
$m$ and  $\Gamma_1 \neq \Gamma_2$, there may be 0, 2 or 4 peaks. In the 
limit of a single semi-infinite
\noindent
chain ($\Gamma_2 \to 0$, $\Gamma_1 \neq 0$),
the two
peaks at $\omega = \pm \omega_2$ merge to produce a central peak 
$(1/2)\delta(\omega)$ representing the contribution of the decoupled $\beta_2$-fermion.
The corresonding contribution to the zero-temperature local susceptibility  is finite:
\bea
 \chi_< (0) = 4 \frac{ K_1 - K_2 }{\Gamma_1 - \Gamma_2}, ~~~
K_a = \theta(|m| - \Gamma_a) \left( \frac{\Gamma_a}{\omega_a} \right)
\left(\frac{|m|-\Gamma_a}{2|m| - \Gamma_a}  \right).\nn
\label{chiK12}\eea
In particular, in the strong localization limit ($|m| \gg \Gamma_{1,2}$)
\be
\chi_< (0)= \sqrt{\frac{2}{|m|}}\frac{1}{\sqrt{\Gamma_1} + \sqrt{\Gamma_2}}.
\label{chi-12-ord-asym}
\ee
In the symmetric case $\Gamma_1 = \Gamma_2$ this transforms to 
\be
\chi_< (0) = \frac{1}{\sqrt{2|m|\Gamma}}. \label{chi-12-ord-sym}
\ee
Like for a single semi-infinite QIC with a boundary 
impurity, the zero-temperature local susceptibility 
is finite,
but its value in the presence of two attached chains is twice smaller than the single-chain value (\ref{chi:m<0}) of $\chi_0$.  It may be interesting to generalize this result to junctions of $n>2$ semi-infinite chains.

\medskip

At $m < 0$ the high-frequency part of the local susceptibility is given by
\bea
\chi_{>} &=& \frac{4}{\pi} \int_{|m|} ^{\infty} \rd \omega~\frac{\sqrt{\omega^2 - m^2}}{\omega^2}
\tanh \left( \frac{\omega}{2T} \right)
\frac{\Gamma_1  (\omega^2 - \Gamma_2 |m|)
 + \Gamma_2  (\omega^2 - \Gamma_1 |m|)}{\left(\omega^2  - \Gamma_1 (2|m| - \Gamma_1)\right)
 \left(\omega^2  - \Gamma_2 (2|m| - \Gamma_2)\right)}.
\label{A-as-m<0}
\eea
This expression can again be rewritten as in Eq.~(\ref{chi>12}). Using that formula, one easily checks that, in the strong localization limit, $|m| \gg \Gamma_1, \Gamma_2$, the high frequency corrections to $\chi_<$ in (\ref{chi-12-ord-asym})
are small, and thus $\chi_0 (0) \simeq \chi_< (0)$.
In the strong hybridization
regime, $|m| \ll \Gamma_1, \Gamma_2$, the local susceptibility $\chi_0$
coincides with $\chi_>$. As follows from our discussion in Sec.~\ref{semi} B3
there the 2-channel Kondo asymptotics of single-chain susceptibilities $\chi_> ^{(1,2)}$
are insensitive to the sign of $m$ ($|m|$ only serves as a low-energy cutoff
of the logarithms). We thus conclude that in the absence of a singular low-temperature Curie susceptibility 
at $m<0$,  $\chi_0$ becomes temperature independent at  
$T \ll {\rm min~}(\Gamma_1, \Gamma_2)$ and is given by (\ref{1=2:const-chi}).
\medskip

Thus, for symmetrically coupled chains (the case of PI),
the susceptibility $\chi_0 (T)$ approaches the constant value (\ref{chi-12-ord-sym})
at any $T \ll |m|$ if $|m| \gg \Gamma$, or the constant value (\ref{cont})
at any $T \ll \Gamma$ if $|m| \ll \Gamma$.

%%%%%%%%%%%%%%%

\subsection{Comparing the local response: boundary impurity versus impurity
in the bulk}

The Kondo-like multiplicative logarithmic renormalization of $\chi_>$ in (\ref{2-kondo-m})
is a feature specific to a single semi-infinite QIC. 
%and, in this sense, is not generic. 
This renormalization emerges in the strong hybridization limit, $|m| \ll \Gamma$,
where it is contributed by the broad
continuum of high-frequency states ($\omega^2 \gg m^2$) forming the "tail" of the spectral weight $A(\omega)$. In this frequency region (and in the leading order in $|m/\omega| \ll 1$)
$A(\omega)$ does not depend on the
mass $m$ and coincides with that for a critical QIC.  According to formulae (\ref{chi>-general:m>0})
or (\ref{chi>-thresh}), the local susceptibility 
and at $T, |m| \ll \Gamma$ displays a logarithmic asymptotics (\ref{m<0;low-T-EK}),
as in the closely related 2-channel Kondo problem.
\medskip

As we have seen %discussed 
in this section, 
for an impurity in the bulk symmetrically coupled %attached 
to two equivalent
chains (which maps to a model of an impurity coupled to a semi-infinite PI chain) the logarithm disappears 
and $\chi(0) = 4/\pi \Gamma$.
This difference stems from the fact that in the topologically massive phases of the two chains
the impurity spin interacts with both boundary MZMs. This suppresses the singularity of the critical low-frequency Green's functions more strongly than in the case of a single chain attached to the impurity.
\medskip

The logarithmic asymptotics (\ref{2H-log}) of the edge susceptibility of a semi-infinite QIC occurs
not only at criticality but also in the quantum critical window at finite $m$, $|m| < T \ll \Gamma$.  The logarithmic multiplicative renormalization of the local
susceptibility is an \emph{unambiguous indication} of the existence of a boundary MZM
at the impurity site.
This is seen at the edge of a QIC, but not at the edge of a PI chain, which only hosts fermionic boundary zero modes.

%%%%%%%%%%%%%%%
\section{Relation to the 1D p-wave superconductor model}\label{1DpwS}

In this section we make contact with the Kitaev's model of a 1D p-wave superconductor \cite{kitaev2}
described by the Hamiltonian (\ref{p-1D-ham}).
The pairing amplitude $\Delta$
is chosen 
to be real and positive. There exists a particle-hole transformation,
$a_n \to (-1)^n a_n^{\dagger}$, that changes the sign of $\mu$ but keeps the rest of the Hamiltonian
(\ref{p-1D-ham})
invariant. Therefore one can always assume that
$\mu \geq 0$. In this region, 
there exists a critical point $\mu = t$ which separates two gapped phases: the topologically
trivial phase at $\mu > t$ and the topologically non-trivial phase\cite{kitaev2, alicea1} at $\mu < t$.
\medskip

Comparing (\ref{ham-a}) and (\ref{p-1D-ham}) one sees that the 1DPS model
%The model (\ref{p-1D-ham}) 
exactly maps onto the QIC %(\ref{ham-a}) 
at $\Delta = t$.
However, universal scaling properties of the two models coincide in the general case
provided that one concentrates on the vicinity of the critical point. Indeed, setting
$\mu = t + m$ 
and introducing the Majorana lattice operators $\zeta_n$ and $\eta_n$
(see Sec.~\ref{reduction}), we first rewrite the Hamiltonian (\ref{p-1D-ham})
in the form
\bea
H =  \frac{\ri t}{4} \sum_n \left(  \eta_n \zeta_{n+1} + \eta_{n+1} \zeta_n - 2 \eta_n \zeta_n\right)
+ \frac{\ri \Delta}{4} \sum_n \left( \eta_n \zeta_{n+1} - \eta_{n+1} \zeta_n  \right)
- \frac{\ri m}{2}  \sum_n \eta_n \zeta_n.
\label{eta-zeta}
\eea
Assuming then that $|m| \ll t$, and passing to the continuum limit according to
the rule 
(\ref{eta-zeta-cont}), we find that the first term on the r.h.s. of (\ref{eta-zeta})
represents a surface term while the remaining part of $H$ transforms to a Hamiltonian
of a massive Majorana fermion given by
\bea
H_M = \int \rd x~{\cal H}_M (x), ~~~
{\cal H}_M (x) = \ri v \eta(x) \p_x \zeta(x) - \ri m \eta(x) \zeta(x),
\label{pw-maj-ham}
\eea
where $v = \Delta a_0$. From this equivalence it follows that close to the Ising criticality
the ordered ($m<0$) and disordered
($m>0$) phases of the QIC adequately describe the topological and non-topological phases
of the 1DPS.
It can be readily seen that in the vicinity of the second Ising critical point,
$\mu = - t + m$ ~($|m|\ll t$) the emerging continuum model still has the Majorana structure
(\ref{pw-maj-ham}) but with $m$ replaced by $-m$. So in this case the nomenclature
of the ordered and disordered phases according to the sign of the mass $m$ is inverted. This is
in agreement with the known fact\cite{kitaev2, alicea1} that the phase located within the interval
$-t<\mu<t$ is topological, whereas the phases occurring at $\mu>t$ and $\mu<-t$
are non-topological.
\medskip

\begin{figure}[hbbp]
\centering
\includegraphics[width=2.9in]{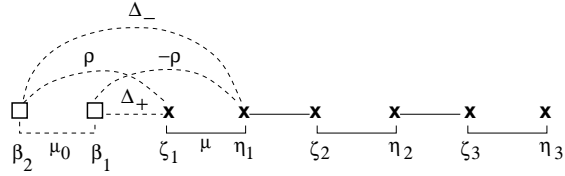}
\caption{\footnotesize Boundary impurity in the Kitaev-Majorana chain
for a 1DPS model at $\mu_0 = 0$.}
\label{pw-imp}
\end{figure}

Now we can construct a model of a 1DPS on a semi-axis $x>0$ with an impurity at the open end.
{Using the spin-fermion equivalence (\ref{spin-fermion-imp})},
we associate the isolated impurity with a local $d$-fermion level,
\[
H_0 = - \mu_0 (d^{\dagger}d - 1/2) = - \ri h_0 \beta_1 \beta_2,
\]
where $h_0 = \mu_0/2$. Accordingly, the local spin susceptibility of the semi-infinite QIC, $\chi_0$,
becomes proportional to the boundary charge susceptibility (local "compressibility")
of the 1DPS, $\kappa_0$:
\[
\chi_0 = 4 \kappa_0, ~~~\kappa_0 = \frac{\p n_d}{\p \mu_0},
\]
where $n_d$ is the mean occupancy of the $d$ level.
The bulk of the system is described by the Majorana Hamiltonian
(\ref{pw-maj-ham}). However, in a 1DPS, {in addition to the standard single-particle
tunneling between the impurity site and the superconductor ($t_0$), the hybridization term may also
contain a local pairing contribution ($\Delta_0$)}:
{
\bea
H'  &=& \frac{t_0}{2} d^{\dagger} a_1 + \frac{\Delta_0}{2} d^{\dagger}a^{\dagger}_1 + h.c.\nn\\
&=& \ri \left( \Delta_+ \beta_1 \zeta_1 +   \Delta_- \beta_2 \eta_1 \right)
+ \ri \rho \left( \beta_2 \zeta_1 - \beta_1  \eta_1\right),
\label{hybrid-PS}
\eea
where
\be
\Delta_{\pm} = \frac{1}{2} \Re e \Delta_0 \pm \frac{t_0}{4}, ~~~
\rho = \frac{1}{2} \Im m \Delta_0. \label{parameters}
\ee
While by a global gauge transformation of the fermion operators $a_n$ it is 
possible to make in Eq.~(\ref{p-1D-ham}) the pairing amplitude $\Delta$  real, 
the phases of the global ($\Delta$) and local ($\Delta_0$) amplitudes
do not generally coincide. This is why
the hybridization term in (\ref{hybrid-PS}) is characterized by three real parameters.}
\medskip

The first term in (\ref{hybrid-PS}) has the same structure
as that already considered in Sec.~\ref{reduction}. 
{To clarify the role of new boundary terms in the Hamiltonian, it is instructive
to consider a toy KM model which involves  two impurity
Majorana sites ($\beta_1$ and $\beta_2$) and two more sites at the open boundary of the chain
($\zeta_1$ and $\eta_1$) -- see Fig.~\ref{pw-imp}.}
{
The Hamiltonian of the model is given by
\be
H_{4} = 
- \frac{\ri \mu_0}{2} \beta_1 \beta_2 - \frac{\ri \mu}{2} \eta_1 \zeta_1 + \ri \Delta_+ \beta_1 \zeta_1 + \ri \Delta_- \beta_2 \eta_1 +  \ri \rho \left( \beta_2 \zeta_1 - \beta_1  \eta_1\right). \label{4site}
\ee
The spectrum of $H_4$ has a $E \to - E$ symmetry and consists of two pairs of levels,
$\pm E_1$ and $\pm E_2$, where
\bea
&&E_1 = \frac{1}{2} \left( \sqrt{\Omega_+} + \sqrt{\Omega_-} \right), ~~~~
E_2 = \frac{1}{2} \left( \sqrt{\Omega_+} - \sqrt{\Omega_-} \right), \label{spec-4}\\
&& \Omega_+ = \left(\frac{\mu + \mu_0}{2} \right)^2 + (\Delta_+ + \Delta_-)^2
+ 4\rho^2, ~~~~
\Omega_- = \left(\frac{\mu - \mu_0}{2} \right)^2 + (\Delta_+ - \Delta_-)^2.
\eea
Assuming that $\Omega_{\pm} \neq 0$, 
it is possible to fine-tune $\mu_0$ to satisfy
the condition $\Omega_+ = \Omega_-$, in  which case the levels $\pm E_2$ become degenerate
at the value $E=0$. This condition translates to}
{
\be
\mu_0 = - \frac{4 (\Delta_+ \Delta_- + \rho^2 )}{\mu}. \label{degen-cond}
\ee
}

The model (\ref{4site}) is expected to yield a satisfactory description of the boundary degrees of freedom in the topologically disordered phase of the 1DPS. In that case it predicts the
degeneracy of the occupied and unoccupied impurity states at the specific value (\ref{degen-cond}) of
$\mu_0$. %proportional to $\Delta_-$. 
At that value of $\mu_0$
the local charge at the impurity
site, $n_d (T=0; \mu_0)$, should display a discontinuity entirely analogous to
the zero-temperature jump of the impurity magnetization in the disordered phase of
the QIC (see Sec.~\ref{dis-phase}). However, the extra $\Delta_-$ and $\rho$-terms in
(\ref{4site}) break the local particle-hole symmetry ($\beta_2 \to - \beta_2$)
and thereby generate a non-zero intrinsic local field at the impurity site. Therefore,
the impurity occupancy $n_d^{\pm}$ right before and after the jump will in general not be situated symmetrically around 1/2, but be shifted by a finite amount.
In the topologically ordered phase, it is still
possible to tune the impurity degrees of freedom to bring about an exact degeneracy between the ground and the first excited states (which are anyway already exponentially close in energy). However,
now they differ  in their topological sector and thus are not connected by a local perturbation. 
This suppresses the singularity in the local compressibility.
\medskip

The above expectations are confirmed by {calculations similar to those done in 
Sec.~\ref{reduction-0}.} As before, the boundary condition for the bulk modes is (\ref{bc-s}). 
In the continuum limit $\zeta_1 \to \sqrt{2a_0} \zeta(a_0)$, $\eta_1 \to \sqrt{2a_0} \eta(a_0)$.
The boundary values of the fields, $\zeta(a_0)$ and $\eta(a_0)$, can be expanded in powers of 
$a_0 \to 0$. {Keeping the leading terms of these expansions 
and using the normal mode expansion (\ref{xi-expan3}) one finds the explicit expressions 
for $\zeta(0)$ and $\eta'(0)$ and thus derives the total Hamiltonian in terms of quasiparticle
operators
\bea
H &=& H_0 + H' \nn\\
&=& - \ri h_0 \beta_1 \beta_2 - \frac{i}{2} \sum_{k>0} \vare_k b_k f_k 
- \ri \lambda_0 \beta_1 \gamma_0 - \ri \tilde{\lambda}_0 \beta_2 \gamma_0 \nn\\
&& - \ri \sqrt{\frac{2}{N}} \sum_{k>0} \lambda_k \beta_1 b_k -
\ri \sqrt{\frac{2}{N}} \sum_{k>0} \tilde{\lambda}_k \beta_2 b_k 
+ \ri \sqrt{\frac{2}{N}} \sum_{k>0} \delta_k \beta_2 f_k -
\ri \sqrt{\frac{2}{N}} \sum_{k>0} \tilde{\delta}_k \beta_1 f_k,
\label{total-hamPS}
\eea
with the coupling constants
\bea
\lambda_0 = 2\Delta_+ \sqrt{\frac{|m|a_0}{v}} \theta(-m), ~~&&~~
\tilde{\lambda}_0 = 2\rho \sqrt{\frac{|m|a_0}{v}} \theta(-m), \nn\\
\lambda_k = \Delta_+ \frac{kv}{\vare_k}, ~~~ \tilde{\lambda}_k = \rho \frac{kv}{\vare_k},
~&&~
\delta_k = \left(\frac{\Delta_- a_0}{v} \right) kv, ~~~
\tilde{\delta}_k = \left(\frac{\rho a_0}{v} \right) kv. 
\eea
Notice that, due to the boundary pairing described by the complex amplitude $\Delta_0$,
both impurity Majorana fermions, $\beta_1$ and $\beta_2$, hybridize with
the boundary zero mode $\gamma_0$, as well as with the gapped continuum states of the bulk spectrum.
}

\medskip

The Greens' functions (GF) of the model (\ref{total-hamPS}) can be easily calculated.
{Since the parameter $v/a_0$ represents an ultraviolet cutoff of the theory, 
small terms of the order $\vare_n a_0/v$ and $|m| a_0/v$ can be systematically dropped. Skipping the details, here we only outline the main results. From the structure of the 
resulting Matsubara GFs $D_{jk}(\vare_n)~(j,k)=1,2$, one can read off that the only effect of finite $\rho$, $\Delta_-$ is to additively renormalize 
the boundary "magnetic field" $h_0$ or, equivalently, the $d$-level local potential $\mu_0$, 
%undergoes an additive renormalization
\be
h_0 ~\to~\tilde{h}_0 = h_0 + \frac{4 a_0}{v} \left( \Delta_+ \Delta_- + \rho^2 \right).
\label{h0-eff}
\ee
This could have been  anticipated qualitatively from Fig.~\ref{pw-imp}:
non-zero values of the couplings $\Delta_-$ and $\rho$ break the local particle-hole  symmetry 
($\beta_2 \to - \beta_2$) 
present at $\mu_0 = 0$
and thus contribute to an effective  pairing of the Majorana fermions $\beta_1$ and $\beta_2$ with an amplitude
$\sim \Delta_+ \Delta_-a_0/v, ~\rho^2 a_0 /v$.
Therefore, the point of the double degeneracy of the spectrum,
i.e., the condition for the boundary impurity degree of
freedom to be classical (that is, conserved by the dynamics), is $\tilde{h}_0 = 0$ 
rather than $h_0= \mu_0 = 0$.

\medskip

{
Using the obtained GFs $D_{jk}(\vare_n)$ and the general formula (\ref{chi-loop-gen})
for the static local susceptibility, 
we find that at the degeneracy point $\tilde{h}_0 = 0$ the local} charge susceptibility of the 1DPS  
is still given by formula (\ref{chi:T=0}), where the width of the $d$-level 
is given by $\Gamma = 4 (a_0/v) (\Delta^2 _+ + \rho^2)$. With this modification and
the redefinition of the spectral degeneracy point ($\tilde{h}_0 = 0$)
the results of the preceding sections fully apply to the 1DPS model.
}

%%%%%%%%%%%%

%%%%%%%%%%%%%%%%%%%%%%%%%%%%%%%%%
\section{Summary and conclusions}\label{summary}

The central result of this paper is summarized in Table~\ref{table:MainIllustration}: The local equilibrium response of an edge or bulk impurity site distinguishes the non-topological and 
%the 
topological phases of the bulk chains, respectively. In the non-topological phase (or, in the disordered phase of the Ising chain), the impurity can be tuned by the local transverse field or the chemical potential through a degeneracy point, where 
{the energy of} a localized boundary mode crosses zero and thus changes occupation in the ground state. At $T=0$ this change is seen as a discontinuity of the transverse magnetization (in 
{the QIC}) %the Ising model) 
or the charge (PI, 1DPS) at the impurity site. For the Ising chain and the PI the degeneracy point is dictated by symmetry to be at $h_0=\epsilon_d=0$, while in a 1DPS one needs to tune the {impurity} potential $\mu_0$ to find the degeneracy point. At finite temperatures, being at the degeneracy point implies a Curie-type divergence of the corresponding susceptibility as $T\to 0$. The coefficient of the Curie law tends to zero upon approaching the topological phase transition. On the topologically (or magnetically) ordered side instead,  we find that it is  impossible to find discontinuous response at the impurity site at $T=0$, and %that 
there is no Curie-like divergence of the susceptibility. In the case of the Ising chain this is simple to understand as the symmetry broken ground state exerts a longitudinal field on the impurity site, which keeps the transverse susceptibility finite despite the occurrence of an exact spectral degeneracy at $h_0=0$. However, it is less obvious to reach the analogous statement for the PI and 1DPS models,
{in which the constituent physical degrees of freedom are spinless fermions and no  symmetry is spontaneously broken}. A unified understanding is possible after a JW transform of the QIC. Then the topological phase of all three models is characterized by boundary {zero modes} 
on the semi-infinite bulk chains. Their coupling with the impurity levels forbids any localized boundary mode to cross zero energy, as a consequence of the level repulsion from the rigid zero mode which cannot be moved by modifying local parameters. 
This avoids the occurrence of any degeneracy not associated with the zero mode itself.
This fact can also be understood as a consequence of the topological protection of the ordered phase against local perturbations: A discontinuity in the local response at $T=0$ would require to switch the parity. Even though at specific values of the parameters $h_0, \epsilon_d$ a global spectral degeneracy is encountered (such as at $h_0= \epsilon_d=0$ for the cases of QIC and PI), this is not reflected in the local susceptibility, because the relevant matrix element connects states of opposite parity. In the topological phase those are exponentially suppressed in the system size. 
\medskip

The smoothness of the local charge response is easily understood in the case of a PI,
upon analyzing the two phases as the impurity potential $\epsilon_d$ is tuned from $+\infty$ to $-\infty$. In the non-topological phase, when the degeneracy point is crossed at $\epsilon_d=0$, an extra charge enters the system and fills a level which is localized close to the impurity. The sudden presence (at $T=0$) of the new charge is seen as a discontinuity in the charge response function $n_d(\epsilon_d)$.
On the topological side, however, a slow decrease of $\epsilon_d$  does not allow an extra charge to enter at the impurity site.
Indeed, at large positive $\epsilon_d$, the level concentrated on the first site of the PI chain is occupied, having a negative energy $\approx -t_0^2/\epsilon_d$, where $t_0$ is the tunneling between the impurity and the PI. As $\epsilon_d $ is reduced, that boundary state  hybridizes  with the impurity site, pushing the energy down to $-t_0$. Meanwhile, the occupancy of the impurity smoothly increases, reaching  $n_d=1/2$ at $\epsilon_d=0$. As $\epsilon_d$ becomes more negative the occupied boundary mode shifts more and more weight onto the impurity site. Note that this level is always occupied, while its weight on the impurity site increases smoothly. 
Nevertheless, for an even PI chain it follows from particle-hole symmetry that at $\epsilon_d=0$ there must be an exact zero mode, also in the topological phase: Indeed, it corresponds to the level localized at the opposite boundary of the PI chain. Its energy changes sign at $\epsilon_d=0$, even though it always remains exponentially close to zero.
The occupation of that boundary mode freely fluctuates at any  finite temperature, while at strictly $T=0$ it undergoes a sharp jump at $\epsilon_d=0$. However, neither of these are visible at the impurity site sitting at the other end of the chain.  
  
\medskip
As usual, the distinction between the topological and the non-topological phase is sharp only at $T=0$, while at finite $T$ a quantum critical window ranging roughly from $-m$ to $m$ smears out the transition between the respective behaviors. Right at criticality, an impurity at the end of a semi-infinite quantum Ising chain realizes the physics of a two-channel Kondo problem, with a logarithmically diverging local susceptibility \cite{EK}. This reflects the fact that one of the two Majoranas that form the impurity spin completely decouples from the rest of the system. In contrast, an impurity embedded in the bulk of a critical chain shows only a saturating susceptibility.

\medskip
We propose to use the difference between the local response in the two phases as an indirect experimental probe for the presence or absence of topologically protected boundary zero modes. 
We expect very similar thermodynamic signatures in the local response of impurities coupled to the edge of large insulators in higher dimensions: in its trivial gapped phase the proximity of such a system will not hinder the discontinuous response of a nearby impurity site~\cite{houzet}. However, in the topological phase hybridization with the gapless edge modes is expected to smoothen out the local response. A closely analogous effect is indeed well-known form coupling quantum dots to gapless Fermi liquids
~\cite{law}. The case of a 1d topological wire is a special case though, in the sense that it does not possess a continuum of edge modes.  
Finally, we expect that this phenomenology is robust to interactions, 
which are currently being discussed in the literature ~\cite{beenak, affleck, schur},
as long as they do not induce a phase transition and gap the edge. {Numerical studies similar to those
of Ref.~\onlinecite{korytar}
might help to obtain a quantitative characterization of interaction effects.}

\medskip

It is worthwhile to compare the equilibrium features in the local response discussed here with probes of Majorana zero modes in transport. The latter focus on the zero-bias anomaly detected in the differential conductance as current is passed {\em through} the wire. Thereby charge enters at one end of the wire and exits at the opposite end. In this {\em transport} set-up the zero mode is constantly populated and emptied. In contrast, the {\em thermodynamics} of the zero mode is essentially blind to the application of local potentials, as it never shifts from zero energy. However, as discussed above, the zero mode has a non-trivial effect on the local spectrum at the impurity, in that it repels the available levels from zero energy. This effect can be probed by transport measurements which use the end of the wire, or an impurity coupled to it, as a quantum point contact between source and drain contacts on either side of the wire. In such a geometry one does not expect the zero mode to contribute to transport directly, not even at zero bias, but to show up in its indirect effects on the levels available for transmission. An analysis of these effects is left for future work.
\medskip

{The results obtained in this paper may be relevant to the studies
of topological effects in junctions and/or quasi-one-dimensional arrays of quantum spin chains and
1D p-wave superconductors. 
However, in this article, we have mostly considered impurities at the end of a semi-infinite chain, or  impurities in the bulk of a single chain. Only  
occasionally we commented on junctions of more than two semi-infinite chains at an impurity site, e.g. when predicting the critical behavior (\ref{chi_n}) of the Curie weight. Our explicit construction of a quasi-local spin operator for junctions of quantum Ising chains suggests that also this case can be solved exactly. It would thus be interesting to investigate potentially non-trivial traces of exchange statistics when comparing the susceptibility of classical spins at junctions of QICs with the analogous response in junctions of 1DPS wires. This question is  of particular interest, given that  such junctions are experimentally relevant elements in any braiding set-up. } 
\bigskip

\medskip

{\sl Acknowledgments:}
The authors express their gratitude to
Boris Altshuler, Michele Fabrizio, Rosario Fazio, Paul Fendley, 
{Leonid Glazman},
Vladimir Kravtsov,  Christopher Mudry, Ady Stern, Andrea Trombettoni, and Alexei Tsvelik for their interest in this work
and stimulating discussions.

%%%%%%%%%%%%
%%%%%%%%%%%%%%%
\appendix
%%%%%%%%%%%%%%%%%

\section{Impurity in a critical QIC. Comparison to the two-channel Kondo problem}
\label{crit}

Consider the model (\ref{H-fin-rep}) and set $m_1 = m_2 = 0$, $v_1 = v_2$.
Passing 
to new %the chirally rotated 
Majorana fields $\xi_{a;R,L} (x)$, defined in (\ref{xi-s}), and using the identification
$
\xi^a _L (-x) = - \xi^a _R (x),
$
one arrives at a model of two chiral (right-moving) Majorana fields on the 
axis $-L < x < L$, coupled to different impurity {Majorana fermions}:
\bea
 H = - \ri h_0 \beta_1 \beta_2 - 2
\ri \sqrt{a_0}\sum_{a=1,2} \tilde{J}_a  \beta_a \xi^a _R (0)
-  \frac{\ri v}{2} \sum_{a=1,2}\int_{-L}^L \rd x~ \xi^a _R (x)
\p_x \xi^a _R (x).
\label{ham-after-map}
\eea
Introducing a single chiral complex field
$
\psi(x) = [\xi^1 _R (x) + \ri \xi^2 _R (x)]/\sqrt{2}
$
and recombining the two impurity Majorana operators $\beta_1$ and $\beta_2$
into a local complex fermionic degree of freedom,
$
d^{\dagger} = (\beta_2 + \ri \beta_1)/2,
$
we
rewrite the Hamiltonian (\ref{ham-after-map}) as follows:
\bea
H &=& \vare_d (d^{\dagger} d - 1/2) -\ri v \int_{-L}^L
\rd x~\psi^{\dagger}(x) \p_x \psi(x)\nn\\
&+& \sqrt{2a_0} \Big\{
(\tilde{J}_1 + \tilde{J}_2) \left[ \psi^{\dagger} (0) d + d^{\dagger} \psi(0) \right]
+  (\tilde{J}_1 - \tilde{J}_2) \left[ \psi (0) d + d^{\dagger} \psi^{\dagger}(0) \right]
\Big\} .\label{H-2ch}
\eea

The two-channel Kondo problem describes two kinds of chiral (right-moving) electrons,
each carrying spin s=1/2, which couple to the impurity spin via two 
channel-dependent,
coupling constants.  
Bosonizing an  XXZ version of this model 
and then refermionizing, it is possible for specially
chosen values of the strength of the couplings that do not flip the impurity spin
(at the so-called Toulouse point)
to map the original Kondo Hamiltonian
to a 
channel anisotropic version of the resonant-level model \cite{EK,book}.
This model has the structure
of the Hamiltonian (\ref{H-2ch})
in which $\tilde{J}_1 = \alpha (g_1 + g_2)$, $\tilde{J}_2 = \alpha (g_1 - g_2)$, where $g_{1,2}$ are the  coupling constants of spin-flip processes associated with the two channels, and $\alpha$ is a constant.
\medskip

So, for a critical QIC with an impurity in the bulk,
the channel-symmetric Majorana version of the resonant-level model 
emerges either at $\tilde{J}_2 = 0$
or $\tilde{J}_1 = 0$. These are the cases when the impurity spin couples either to the right
semi-axis or to left semi-axis only. In other words, only a semi-infinite QIC with
a boundary impurity spin exhibits the Majorana resonant-level behavior
typical for the channel-symmetric two-channel Kondo problem. All cases with
$\tilde{J}_1 \neq \tilde{J}_2$ map to channel-asymmetric two-channel Kondo problems. 
In the special case when
the impurity couples to its right and left nearest-neighbor spins symmetrically ($\tilde{J}_1
= \pm \tilde{J}_2$), the mapping is to a standard, one-channel resonant-level model.

%%%%%%%%%%%%%%%%%%%%%%%%%%%%%%%%%%%%%%%%%%%%%%%%%%
\section{Diagonalization of the massive Majorana model on a semi-axis}\label{diagonalization}
Consider  the Hamiltonian
of a massive Majorana fermion: %defined on a semi-axis:
\bea
&& H_M =  \frac{1}{2}\int_0 ^L \rd x~\xi^T (x) \hat{h}(x) \xi (x),
\label{H-Maj}\\
&&\xi (x) = \left(
\begin{array}{clcr}
\xi_R (x)\\
\xi_L (x)
\end{array}
\right), ~~\hat{h}(x) = - \ri v \p_x \hat{\s}_3 + m \hat{\s}_2.\label{Maj-spinor}
\eea
Diagonalization of this model is standard (see e.g. Ref.~\onlinecite{lo}). 
%Below we will closely follow Ref. \cite{lo}.
We are looking for solutions of
the Dirac equation on an
interval $0<x<L$, assuming that $L \to \infty$:
\be
\hat{h} (x) \chi_{\vare}(x) = \vare \chi_{\vare}(x), ~~~~\chi_{\vare}(x) = \left(
\begin{array}{clcr}
u_{\vare}(x)\\
v_{\vare}(x)
\end{array}
\right).
\label{eigen}
\ee
The boundary condition
\be
u_{\vare} (0) = - v_{\vare}(0)
\ee
follows from (\ref{bc-xi}).
From the property $\hat{h}^*(x) = - \hat{h}(x)$ it follows that
$\chi^* _{\vare} (x) = \chi_{-\vare} (x)$.
Therefore the spectrum consists
of $(\vare, - \vare)$ pairs and, possibly, a zero-energy mode ($\vare = 0$).
This leads to the following %decomposition 
normal-mode expansion of the Majorana field $\xi(x)$:
\bea
\xi(x) = \gamma_0 \chi_0 (x) + \sum_{\vare > 0}
\left[ \gamma_{\vare} \chi_{\vare} (x) +
\gamma^{\dagger}_{\vare} \chi^* _{\vare} (x) \right] = \xi^{\dagger} (x).
\label{xi-expan2}
\eea
Here $\gamma^{\dagger}_0 = \gamma_0$ is a Majorana operator describing the localized zero mode
with a normalizable wave function $\chi_0 (x)$, and
$\gamma_{\vare}, ~\gamma^{\dagger}_{\vare}$ are standard second-quantized fermionic operators
describing the states within the continuous part of the spectrum and satisfying the standard algebra
\bea
\{ \gamma_{\vare}, \gamma^{\dagger}_{\vare'} \} = \delta_{\vare\vare'},~~~
\{\gamma_{\vare}, \gamma_0\} = 0, ~~~\{ \gamma_0, \gamma_0 \} = 2\gamma^2 _0 =1.
\label{gamma-algeb}
\eea
Notice that in (\ref{xi-expan2})
the summation in the second term goes over  states of positive energy only. 
Substituting the expansion (\ref{xi-expan2}) into (\ref{H-Maj}) we arrive at the diagonalized Hamiltonian
\be
H = \sum_{\vare> 0} \vare \hat{\gamma}_{\vare} \hat{\gamma}_{\vare}, ~~~~
[H,\gamma_0] = 0,
\label{HM-diagon}
\ee
which is valid if the eigenvectors belonging to the continuous part of the spectrum satisfy the orthonormalization conditions
\bea
&&\int_{0}^L \rd x~\left[ u^* _{\vare} (x) u_{\vare'} (x) +   v^* _{\vare} (x) v_{\vare'} (x) \right]
= \delta_{\vare\vare'}, ~~~% \label{ortho1}\\
\nn\\
&&\int_{0}^L \rd x~\left[ u_{\vare} (x) u_{\vare'} (x) +   v_{\vare} (x) v_{\vare'} (x) \right]
= 0.
\label{ortho}
\eea
The algebra (\ref{xi-algebra}) of the Majorana fields implies the completeness relations:
\begin{flushleft}
\bea
&&u_0 (x) u_0 (x') + \sum_{\vare>0}
\left[ u_{\vare} (x) u^* _{\vare} (x') + c.c. \right] = \delta(x-x'),~~~~~~~ \label{compl1}\\
&&v_0 (x) v_0 (x') + \sum_{\vare>0}
\left[ v_{\vare} (x) v^* _{\vare} (x') + c.c. \right] = \delta(x-x'),~~~~~~~\label{compl2}\\
&&u_0 (x) v_0 (x') + \sum_{\vare>0}
\left[ u_{\vare} (x) v^* _{\vare} (x') + c.c. \right] = 0. ~~~~~~~\label{compl3}
\eea
\end{flushleft} 
The
solution of the Dirac equation (\ref{eigen}) has the following form.
The 
normalizable zero-energy solution only exists
for $m<0$: %(see Appendix \ref{zero-mode}):
\bea
&& \chi_{0}(x) =  \left( 
\begin{array}{clcr}
~1\\
-1
\end{array}
\right) \phi_0 (x), \nn\\
&&\phi_0 (x) = \sqrt{\frac{|m|}{v}} ~\theta(-m) \exp \left( - |m| x/v \right),  
\label{spinor-0mode-fin}\\
&& \la  \chi_0 | \chi_0 \ra = \int_{0}^{\infty} \rd x~ \left[  u^2 _0 (x) + v^2 _0 (x) \right]
= 1. \nn%\label{0mode-norm}
\eea
This is an indication of the topological nature of the
ordered ($m<0$) massive phase of the QIC. There also exists a continuum
of extended states with energies $\vare^2 \geq m^2$. %According to (\ref{xi-expan2}),
The latter  
are parametrized by the
quantum numbers $k_j = \pi j/L > 0 ~(j=1,2, \ldots , N=L/a_0)$:
\bea
&&\chi_{k} (x) \equiv 
\left[
\begin{array}{clcr}
u_k (x) \nn\\
v_k(x)
\end{array}
\right]
= 
\left[
\begin{array}{clcr}
W_k (x) \nn\\
- W^* _k(x)
\end{array}
\right], ~~~~\vare_k = \sqrt{k^2 v^2 + m^2},
\label{chi-fin1}\\
&&
W_k (x) =
\frac{1}{\sqrt{2L}} \left[
\cos (kx - \theta_k) + \ri \sin kx \right], ~~~k>0.
\label{W}
\eea
The phase shift $\theta_k$ is defined through the relation $\tan \theta_k =m/kv$.
Using (\ref{spinor-0mode-fin}) and (\ref{W}) one can easily verify the relations
(\ref{ortho}) and (\ref{compl1})--(\ref{compl3}).
The expansion (\ref{xi-expan2}) takes its final form
\bea
\xi(x) = \gamma_0 \chi_0 (x) + \sum_{k > 0}
\left[ \gamma_{k} \chi_{k} (x) +
\gamma^{\dagger}_{k} \chi^* _{k} (x) \right],
\label{xi-expan3}
\eea
and the diagonalized Hamiltonian becomes
\bea
H = \sum_{k>0} \vare_k \hat{\gamma}^{\dagger}_{k} \hat{\gamma}_{k}
+ {\rm const}.
\label{H-diag}
\eea

%%%%%%%%%%%%%%%%%%%%%%%
\section{Majorana Green's functions for the impurity}\label{derivGF}

Here we derive  
the $d$-fermion GF $G(\vare_n)$ in the the most general case described by the
Hamiltonian (\ref{maj-ham-final}).
$G(\vare_n)$ can be expressed in terms of the impurity Majorana GFs $D_{jk} (\vare_n)$:
\bea
G(\vare_n) = \frac{1}{4} \left[ D_{11} (\vare_n) +
D_{22} (\vare_n) + 2\ri D_{21} (\vare_n) \right], \label{G-D}\\
\eea
where
\bea
D_{jk} (\vare_n) = - \int_0 ^{1/T} \rd \tau~e^{i\vare_n \tau}
\la T_{\tau} \beta_j (\tau) \beta_k \ra. \label{Djk}
\eea
Using the equations of motion for the Heisenberg
operators $\beta_j (\tau)$ we obtain a set of  equations for the GFs:
\bea
\ri \vare_n D_{11}(\vare_n)
&=& 2 - 2\ri h_0 D_{21} (\vare_n)
- 2\ri \lambda_{10} L_1 (\vare_n)
- 2\ri  \sqrt{2/N}\sum_{k>0} \lambda _{1k} B_{1k} (\vare_n);
\label{basic-11}\\
\ri \vare_n D_{21} (\vare_n)
&=&  2\ri h_0 D_{11}(\vare_n)
- 2\ri \lambda_{20} L_2 (\vare_n)
- 2\ri \sqrt{2/N}\sum_{k>0} \lambda _{2k} B_{2k} (\vare_n);
\label{basic-21}\\
\ri \vare_n ~L_a (\vare_n)
&=& \ri \lambda_{a0} D_{a1}(\vare_n);
\label{L-bet}\\
\ri \vare_n B_{ak}(\vare_n)
&=& - \ri \vare_{ak} F_{ak} (\vare_n)
+ 2\ri \sqrt{2/N}\lambda_{ak} D_{a1} (\vare_n);
\label{B-bet}\\
\ri \vare_n F_{ak} (\vare_n)
&=& \ri \vare_{ak} B_{ak} (\vare_n) .
\label{F-bet}
\eea
In the above formulas 
the following notations %for the Fourier transforms of all GFs
have been used:
\bea
L_a (\vare_n) &=&  - \int_0 ^{1/T} \rd \tau~ e^{i\vare_n \tau}
\la T_{\tau} \gamma_{a0}(\tau) \beta_1(0)  \ra, \nn\\
B_{ak} (\vare_n) &=&  - \int_0 ^{1/T} \rd \tau~ e^{i\vare_n \tau}
\la T_{\tau} b_{ak}(\tau) \beta_1(0)  \ra, \nn\\
F_{ak} (\vare_n) &=&  - \int_0 ^{1/T} \rd \tau~ e^{i\vare_n \tau}
\la T_{\tau} f_{ak}(\tau) \beta_1(0)  \ra,  ~~~~~(a=1,2).\nn
\eea
From Eqs.(\ref{basic-11})--(\ref{F-bet}) one obtains a closed set of equations
for the Majorana GFs $D_{jk}(\vare_n)$:
\bea
&& \Omega_1 (\vare_n)
D_{11}(\vare_n) = 2 - 2\ri h_0  D_{21}(\vare_n), \nn\\
&& \Omega_{2}(\vare_n)
D_{21}(\vare_n) = 2\ri h_0  D_{11}(\vare_n),
\nn
\eea
where
\[
\Omega_a (\vare_n) = \ri \vare_n  - \frac{2 \lambda_{a0}^2}{\ri \vare_n}
+ \ri \vare_n \frac{8}{N} \sum_{k>0} \frac{\lambda^2 _{ak}}{\vare^2 _n + \vare^2 _{1k}}.
\]
The sums over $k>0$ are easily evaluated:
\bea
\frac{1}{N} \sum_{k>0} \frac{\lambda^2 _{ak}}{\vare^2 _n + \vare^2 _{ak}}
= \frac{\tilde{J}^2 _a a_0}{\pi} \int_0^{\infty} 
\rd k~\frac{k^2v^2_a}{\vare^2_{ak} (\vare^2 _{ak} + \vare^2 _n)}
= \left(\frac{\tilde{J}^2 _{a} a_0}{2 v_a} \right)
 \frac{\sqrt{\vare^2 _n + m^2 _a} - |m_a|}{\vare^2 _n}.  
\label{eval-int}
\eea
Denoting
\bea
&&\Gamma_a = \frac{4 \tilde{J}^2 _a a_0}{v_a}, 
\label{gamma-defin}\\
&&\Delta_a(\vare_n) = \vare^2 _n 
 + \Gamma_a \left( \sqrt{\vare^2 _n + m^2 _a} - m_a \right), 
\label{Delta=denom} 
\eea
we arrive at the final expressions for the impurity Majorana GFs,
\bea
&&D_{11} (\vare_n) = - \frac{2\ri \vare_n \Delta_2 (\vare_n)}
{\Delta_1 (\vare_n) \Delta_2 (\vare_n) + 4 h^2 _0 \vare^2 _n}, 
\label{D11-fin}\\
&& D_{21} (\vare_n) = 
- \frac{4\ri h_0 \vare^2 _n}{\Delta_1 (\vare_n) \Delta_2 (\vare_n) + 4 h^2 _0 \vare^2 _n}.
\label{D11-21-fin}
\eea
By symmetry, the remaining GFs are
\bea
&&D_{22} (\vare_n) = - \frac{2\ri \vare_n \Delta_1 (\vare_n))}
{\Delta_1 (\vare_n) \Delta_2 (\vare_n) + 4 h^2 _0 \vare^2 _n},
\label{D22-fin}\\
&&D_{12} (\vare_n) = - D_{21} (\vare_n). \label{D22-12-fin}
\eea
According to (\ref{G-D}), the local GF of the complex fermion, $G(\vare_n)$,
is given by
\bea
G(\vare_n) &=& -\frac{1}{2} \frac{\ri \vare_n [\Delta_1 (\vare_n) +\Delta_2 (\vare_n) + 4\ri h_0 \vare_n]}
{\Delta_1 (\vare_n) \Delta_2 (\vare_n) + 4 h^2 _0 \vare^2 _n}.
\label{G-G1}
\eea
In particular, at $\Gamma_2 = 0$,
\bea
G(\vare_n) = \frac{1}{2\ri \vare_n} \left[ 1 -
\frac{(\ri \vare_n - 2h_0)^2}{\Delta_1 (\vare_n) + 4h^2 _0} \right].
\label{G1}
\eea
At $h_0 = 0$ the expressions of all impurity GFs simplify:
\bea
&&D_{aa}(\vare) = \frac{2\ri \vare_n}{\Delta_a (\vare_n)},~~~~
D_{12} (\vare_n) = D_{21} (\vare_n) = 0, \label{D-h=0}\\
&& G(\vare_n) = \frac{1}{4} \left[ D_{11} (\vare_n) + D_{22} (\vare_n) \right]
=
- \frac{\ri \vare_n}{2} \left[ \frac{1}{\Delta_1 (\vare_n)} +
 \frac{1}{\Delta_2 (\vare_n)}\right].\label{G-D2}
\eea

When both Majorana chains (channels) are identical 
($v_1 = v_2 \equiv v, ~m_1 = m_2 \equiv m$)
and the hybridization constants also coincide ($\Gamma_1 = \Gamma_2 \equiv \Gamma$),
we get:
\bea
G(\vare_n) = -\frac{\ri \vare_n}{\Delta(\vare_n) - 2\ri h_0 \vare_n}
= \frac{\ri \vare_n}{\ri \vare_n (\ri \vare_n + 2h_0) - \Gamma
(\sqrt{\vare^2 _n + m^2} - m)}.
~~~\label{G-sym}
\eea
Under the replacements $2h_0 \to - \vare_d$ $G(\vare_n)$ in (\ref{G-sym})
coincides with the GF of the impurity $d$-fermion in a slightly non-critical semi-infinite PI
chain (see Appendix \ref{peierls}).

%%%%%%%%

\section{Relation to Peierls insulator} \label{peierls}

In this Appendix we consider the two-chain massive Majorana model (\ref{H-fin-rep-xi}) in the case
when all its parameters referring to different chains coincide:
\[
\tilde{J}_a = \tilde{J}, ~~~v_a = v, ~~~m_a = m ~~~~(a=1,2).
\]
In this case the model acquires an extra O(2)-symmetry related to global rotations
of the Majorana vector $\vxi = (\xi_1, \xi_2)$. Therefore it is natural to reformulate the problem
in terms of a complex Dirac field,
\be
\psi_{\nu} (x) = \frac{\xi_{1\nu} (x) + \ri \xi_{2\nu} (x)}{\sqrt{2}},
\label{xi12-psi}
\ee
$\nu = R,L$ being the fermion chirality index.
Passing simultaneously from the pair of boundary Majorana operators ($\beta_1, \beta_2$)
to the second quantized operators of the complex $d$-fermion,~
$
d^{\dagger} = (\beta_2 + \ri \beta_1)/2,
$
~we transform the Hamiltonian (\ref{H-fin-rep-xi}) to a semi-infinite model of a \emph{massive},
U(1)-symmetric
resonant-level model:
\bea
H  &=& \vare_d (d^{\dagger} d - 1/2) - 2\sqrt{2a_0}~\tilde{J} \left[
d^{\dagger} \psi_R (0) + \psi^{\dagger}_R (0) d
\right]\nn\\
&+& \int_0^L \rd x~ \Big[
\ri v_F \left(
\psi^{\dagger}_L (x) \p_x \psi_L (x) - \psi^{\dagger}_R (x) \p_x \psi_R (x)
\right) - \ri m \psi_R (x) \psi_L (x)
\Big].
\label{H-mass-u1}
\eea
Here $\vare_d = - 2h_0$, and the fermionic field satisfies the
boundary condition $\psi_R (0) = - \psi_L (0)$. 
\vskip 0.2 truecm

It is well-known that the Lorentz-invariant Dirac model (\ref{H-mass-u1}) emerges in
the continuum description of a spinless version of a Peierls insulator at half filling
(the so-called polyacetelyne model)\cite{takayama, ssh}. In a broken-symmetry, spontaneously dimerized state the fermionic part
of the PI Hamiltonian is a tight-binding model with alternating
nearest-neighbor hopping amplitudes:
\bea
H_{\rm PI} = - \sum_{n\geq 1} t_{n,n+1}\left(  c^{\dagger}_n c_{n+1} + h.c.\right),~~~~
t_{n,n+1} = t - (-1)^n \Delta. \label{pi}
\eea
At $|\Delta| \ll t$ a continuum limit can be taken in (\ref{pi}),
\[
c_n \to \sqrt{a_0}~ \left[ \ri^n \psi_R (x) + (-\ri)^n \psi_L (x) \right],
~~~~(k_F = \pi/2a_0),
\]
yielding the bulk term
in (\ref{H-mass-u1}) with $v_F = 2ta_0$ and $m = 2\Delta$.
The ground state of the Hamiltonan (\ref{pi}) is dimerized.  %and $\mathbb{Z}_2$-degenerate. 
Accordingly,
there are two massive phases, $\Delta = \pm \Delta_0$, separated by a gapless metallic state ($\Delta = 0)$. 
The two phases with the same $|\Delta|$ have identical bulk spectra. 
Their topological difference \cite{shen} shows up in the boundary conditions at the edges of a finite
chain. Repeating the Kitaev's argument and turning to special cases $\Delta = \pm t$
(with $t>0$) one finds a topologically degenerate ground state with
two boundary zero modes at $\Delta < 0$ and a nondegenerate ground state at $\Delta > 0$.
These zero modes are bound states of a massive complex fermion, each state carrying a fractional
fermion number (charge) $q_F = 1/2$ \cite{ssh, niemi}.
\medskip

The Green's function of the impurity $d$-fermion for the PI,
$G(\vare_n)$, is given in Appendix \ref{derivGF}, Eq.~(\ref{G-sym}). Passing to the retarded GF
${\cal G} (\omega + \ri \delta)$ we can calculate the spectral weight $A_{\rm f}(\omega)$ of the $d$-electron
states. The result is given by Eqs.~(\ref{A-m>0}) and (\ref{A-m<0}).
\medskip

%%%%%%%%%%%%

%\newpage
%%%%%%%%%%%%%%

%%%%%%%

\end{document}